\documentclass[12pt]{article}
\usepackage{graphics,epsfig,cite,latexsym}

\newcommand{\BC}{\begin{center}}    \newcommand{\EC}{\end{center}}
\newcommand{\BE}{\begin{equation}}  \newcommand{\EE}{\end{equation}}
\newcommand{\BA}{\begin{eqnarray}}  \newcommand{\EA}{\end{eqnarray}}
\newcommand{\BN}{\begin{enumerate}} \newcommand{\EN}{\end{enumerate}}
\newcommand{\BI}{\begin{itemize}}   \newcommand{\EI}{\end{itemize}}
\newcommand{\BFig}{\begin{figure}}  \newcommand{\EFig}{\end{figure}}
\newcommand{\BTab}{\begin{table}}   \newcommand{\ETab}{\end{table}}
\newcommand{\BT}{\begin{tabular}}   \newcommand{\ET}{\end{tabular}}
\newcommand{\BM}{\begin{minipage}}  \newcommand{\EM}{\end{minipage}}
\newcommand{\I}{\item}

\newcommand{\LL}{\label}
\newcommand{\NN}{\nonumber}

\def\JL#1#2#3#4{{#1} {\bf #2} (#3) #4}

\def\MBF#1{\mbox{\bf #1}}

\newcommand{\pb}{\mbox{$\mathrm{\bar{p}}$}}
\newcommand{\ppb}{\mbox{$\mathrm{\bar{p}p}$}}
\newcommand{\Mppb}{\mbox{$m_{\mathrm{\bar{p}p}}$}}
\newcommand{\Mp}{\mbox{$m_{\mathrm p}$}}
\newcommand{\GeV}{\mbox{\rm GeV}}        
\newcommand{\GeVc}{\mbox{\rm GeV/c}}        
\newcommand{\MeV}{\mbox{\rm MeV}}        
\newcommand{\MeVc}{\mbox{\rm MeV/c}}        
\newcommand{\RHO}{\varrho}       

\newcommand{\Mooa}{M_{a}}
\newcommand{\Moob}{M_{b}}

\newcommand{\Iref}[1]{$^{#1}$}

\newcommand{\Instfoot}[2]{\small $^{#1)}$~{\it #2}\\}  

\begin{document}

\thispagestyle{empty}
\begin{flushright}
February 18, 2002
\end{flushright}

\newcommand{\iMuen}{1}
\newcommand{\iPSI}{2}
\newcommand{\iUNIZH}{3}
\newcommand{\iETHZ}{4}

\begin{center}
{\Large\sf Pion Correlations and Resonance Effects in  
           $\ppb$ Annihilation to $4\pi^0$ at Rest}
\\[2\baselineskip]

{\sc
O.~Kortner\Iref{\iMuen}, 
M.P.~Locher\Iref{\iPSI,\iUNIZH}, 
V.E.~Markushin\Iref{\iPSI}, 
P.~Weber\Iref{\iETHZ}, 
O.~Wigger\Iref{\iPSI,\iUNIZH} 
}
\end{center}

\noindent%
\Instfoot{\iMuen}{Ludwig--Maximilians--Universit\"at, M\"unchen}
\Instfoot{\iPSI}{Paul Scherrer Institut (PSI), Villigen, Switzerland}
\Instfoot{\iUNIZH}{University of Z\"urich, Switzerland} 
\Instfoot{\iETHZ}{ETH-IPP Z\"urich, Switzerland} 


\vspace{1\baselineskip}

\begin{abstract}

We study $\pi^0\pi^0$ correlations in the exclusive reaction $\ppb\to4\pi^0$ 
at rest with complete reconstruction of the kinematics for each event.  
The inclusive distribution shows a dip at small invariant mass of the pion pair 
while a small enhancement in the double differential distribution is observed
for small invariant masses of both pion pairs.
Dynamical models with resonances in the final state are shown to be consistent 
with the data while the stochastic HBT mechanism is not supported 
by the present findings. 
 
\end{abstract}


\section{Introduction} \LL{Intro}
 
  Nucleon-antinucleon annihilation into multi-pion states offers the 
possibility of studying Bose-Einstein (BE) correlations under controlled 
conditions.  
While the study of BE correlations in {\it inclusive} distributions in 
$N\bar{N}$ annihilation in the conventional Hanbury-Brown--Twiss (HBT) 
framework \cite{HBT} has a long history \cite{Go60,CO74,KP73}, 
the use of {\it exclusive} distributions is relatively new.  
The first steps in this direction were  
made in \cite{CPLEAR97,CPLEAR98} where the exclusive reactions 
$\ppb\to 2\pi^+2\pi^-$ and $\ppb\to 2\pi^+2\pi^-\pi^0$
at rest were studied on the basis of minimum bias CPLEAR data. 
  For these data, the complete kinematical reconstruction of each event
allows the direct determination of the square of the reaction amplitude 
and the dynamics of the BE correlations is not obscured by integration 
over spectators and the usage of conventional reference samples. 

  The present paper extends the analysis of the exclusive distributions   
for the $\ppb$ annihilation at rest to the case of the $4\pi^0$ final state 
on the basis of the Crystal Barrel (CB) data.  
An appealing property of the $4\pi^0$ channel is that there are no  
$\rho$ mesons, which strongly affect the final states with charged pions.  
Therefore the comparison of the $\pi^0\pi^0$ correlations with the 
charged pion correlations can clarify the origin of the signals observed. 

  The HBT mechanism \cite{HBT}, which is based on pion emission with stochastic 
phases over an extended emission volume, predicts an {\it enhancement} 
of like sign pion pairs at low relative momentum $Q^2$.  
This kind of enhancement seen for $\pi^+\pi^+$ and $\pi^-\pi^-$ pairs in 
many reactions is usually called a BE signal.   
For the first time in an annihilation 
reaction, the {\it inclusive} $2\pi^0$ pair distribution for $\ppb\to 4\pi^0$ 
shows no such enhancement contrary to the case of $\ppb\to 2\pi^+2\pi^-$ 
\cite{CPLEAR97} where a weak enhancement was seen in the corresponding 
$\pi^+\pi^+$ and $\pi^-\pi^-$ distributions.  In the double differential 
distribution, which is a more sensitive observable, a small enhancement is 
seen in the kinematical region where the $4\pi^0$ system forms two pion pairs,
both having small invariant masses.
  As we shall discuss, the strength and the shape of the signal for
the differential correlations do not favor an interpretation 
in terms of the conventional HBT picture of BE correlations.  
This conclusion is based on simulations of resonance
production in the $4\pi^0$ final state which qualitatively explain   
the observed effects. 
   
  The plan of the paper is as follows.  
In section~\ref{DataAnal} we describe the analysis of the data and the results 
for the single variable distributions.  
In section~\ref{METH4pi0}, the formalism of double differential distributions
is recapitulated.     
Section~\ref{DISC} presents detailed model calculations for the dominant
resonance mechanisms which are compared to the data.  
Partial projections similar to the analysis performed earlier in the
$2\pi^+2\pi^-$ case are shown as well and confronted with our dynamical model.
Section~\ref{CONCL} gives a summary and conclusions.

\section{Analysis of the $4\pi^0$ data}
\label{DataAnal}

The Crystal Barrel experiment \cite{CBdetector,CB} at the Low Energy Antiproton
Ring (LEAR) at CERN was conceived for high statistics meson spectroscopy.
The setup relevant for the present study 
used a cylindrical detector consisting of a hydrogen gas target, 
tracking chambers for charged particles (proportional 
and jet-drift-chambers), and an electromagnetic calorimeter with 1380
CsI crystals of 15 radiation lengths each for the detection of photons.  
The calorimeter is surrounded by a 1.5~T solenoid magnet.

\subsection {Event selection}

The $4\pi^0$ data analyzed in this study was taken by Crystal Barrel in 
November 1994 using antiprotons stopped in the hydrogen gas target at 12~bar 
with the ``all-neutral trigger'', i.e.\ only events
with no signal in the proportional chambers were recorded.
Offline, the data was subjected to the following selection criteria:
\BI
\I exactly 8 photon energy deposits (PED) in the calorimeter;
\I no PED adjacent to the beam pipe (risk of missing energy);
\I no $\pb$ pileup condition;
\I no charged tracks in proportional chambers or jet-drift-chambers
   (eliminates backscatters, trigger inefficiency, and pair production);
\I no PED below 20 \MeV (suppresses split-offs);
\I total momentum between 0 and 200~\MeVc,
   and total energy $\Mppb\pm 173\;$\MeV. 
\EI
For this set of events, a kinematic fit is performed and the most likely
photon combinations are chosen.
Combinatorial background is determined by simulation to be
of the order of 18 percent.
The photon energy resolution is $\sigma_E/E = 2.4\%/\sqrt[4]{E/\GeV}$.


\begin{figure}[htbp]
\centering 
\mbox{\epsfysize=8cm \epsffile{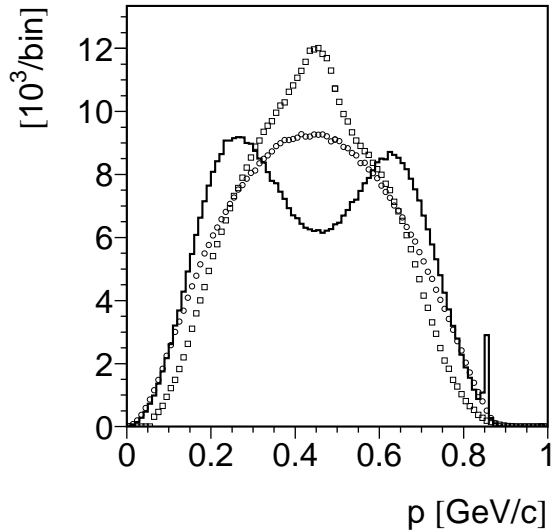}}
\caption{\label{p0incl}
The measured single-pion momentum distribution $dN_{\pi}/d|\MBF{p}_{\pi}|$
for the $4\pi^0$ channel (thick histogram).
The pure phase space distribution is shown with $\protect\circ$ and  
the single-pion momentum distribution for the 
$2\pi^+2\pi^-$ channel \protect\cite{CPLEAR97} with $\protect\Box$.}    
\end{figure}

The measured single pion inclusive momentum distribution is plotted in 
Fig.\ref{p0incl}. 
  Contrary to the case of four charged pions $2\pi^+2\pi^-$ \cite{CPLEAR97}, 
the shape of the $\pi^0$ momentum distribution differs significantly from
phase space due to the strong production of resonances in the mass region 
between 1.2~\GeV\ and 1.7~\GeV\ as discussed in \cite{Ko98}.   
The broad peak at 0.2 \GeVc\ is due to pions recoiling against a heavy
resonance like $a_2(1660)$ or $\pi_2(1670)$,
while the maximum around 0.65 \GeVc\ stems from the decay of 
lighter resonances like $f_2(1270)$ and maybe $\pi(1300)$. 
The narrow peak at $0.85\;\GeVc$ corresponds to the 
$\ppb\to\pi^0+\eta_{\to3\pi^0}$ decay.   
We shall discuss the resonance production in detail in section~\ref{DISC}.

\subsection{Correlation functions for inclusive distributions}
\label{CF}

   For the benefit of the reader, we summarize below the standard BE formalism 
according to \cite{CPLEAR97}.    
   The single-particle inclusive density $\rho_1(p_1)$ and the two-particle
inclusive density $\rho_2(p_1,p_2)$ are related to the differential
cross-sections by 
\begin{eqnarray}
   \rho_1(p_1)     & = & \sigma^{-1} \frac{d\sigma}{d^3\MBF{p}_1/(2E_1)} \\ 
   \rho_2(p_1,p_2) & = & \sigma^{-1} 
             \frac{d\sigma}{d^3\MBF{p}_1/(2E_1)\; d^3\MBF{p}_2/(2E_2)} \ \ .
\end{eqnarray} 
   One of the definitions of pion pair correlations is based on the
formula 
\begin{eqnarray}
   c(p_1,p_2) = \rho_2(p_1,p_2) - \rho_1(p_1) \rho_1(p_2) \quad . 
\label{cf}
\end{eqnarray}
Alternatively the two-particle correlations can be described in terms of
the ratio
\begin{eqnarray}
   C(p_1,p_2) = \frac{\rho_2(p_1,p_2)}{\rho_0(p_1,p_2)} \quad ,
\label{C}
\end{eqnarray}
where $\rho_0(p_1,p_2)$ is the two-particle distribution {\it in the absence
of correlations}, with various prescriptions being used in the literature. 
A choice consistent with Eq.(\ref{cf}) is the product of the 
single-particle densities $\rho_0(p_1,p_2) = \rho_1(p_1) \rho_1(p_2)$.

  Averaging over angles and momenta gives a correlation function
depending on one variable, the two-pion invariant mass $M$:
\BA
   C(M) & = & \frac{\rho_2(M)}{(\rho_1\cdot\rho_1)(M)} \LL{CM} 
\\
   \rho_2(M) & = & \int \delta\left(M - \sqrt{(p_1+p_2)^2}\right)
   \rho_2(p_1,p_2) \frac{d^3\MBF{p}_1 d^3\MBF{p}_2}{(2E_1)(2E_2)} \LL{rho2} 
\\
   (\rho_1\cdot\rho_1)(M) & = & \int \delta\left(M - \sqrt{(p_1+p_2)^2}\right)
   \rho_1(p_1) \rho_1(p_2) \frac{d^3\MBF{p}_1 d^3\MBF{p}_2}{(2E_1)(2E_2)} 
   \LL{rho11}
\EA
The invariant mass $M$ is uniquely related to the square of the 
momentum difference:
\BE
   (p_1 - p_2)^2 = 4\mu^2 - M^2 = -Q^2,
\LL{MQ}
\EE
where $\mu$ is the pion mass and $\MBF{Q}$ is the difference of
the three-momenta of the two pions in their center-of-mass system (CMS), 
therefore the variables $M^2$ and $Q^2$ are equivalent.

Because of the total energy-momentum conservation, the ratio $C(M)$ is not a
constant even if the distributions $d\sigma/(d^3\MBF{p}_1/2E_1)$ and
$d\sigma/(d^3\MBF{p}_1/2E_1)(d^3\MBF{p}_2/2E_2)$ are determined by phase
space alone (see \cite{CPLEAR97,CPLEAR98}).  
This effect becomes negligible for reactions at much higher energy, 
but it is important for annihilation at rest.

\subsection{Single-variable two-pion correlations}
\LL{InclCor}
  
In this subsection we present the single-variable two-pion correlation  
$C(M)$ which has been frequently used in previous analyses.
In order to isolate the correlation effects we compare the experimental
density with a four-pion phase space distribution corrected for experimental
cuts and efficiencies in the same way as the data.

\begin{figure}[htbp]
\centering
\mbox{\hspace*{-10mm}
  \parbox{7cm}{\center(a)\\ \mbox{\epsfysize=7cm\epsffile{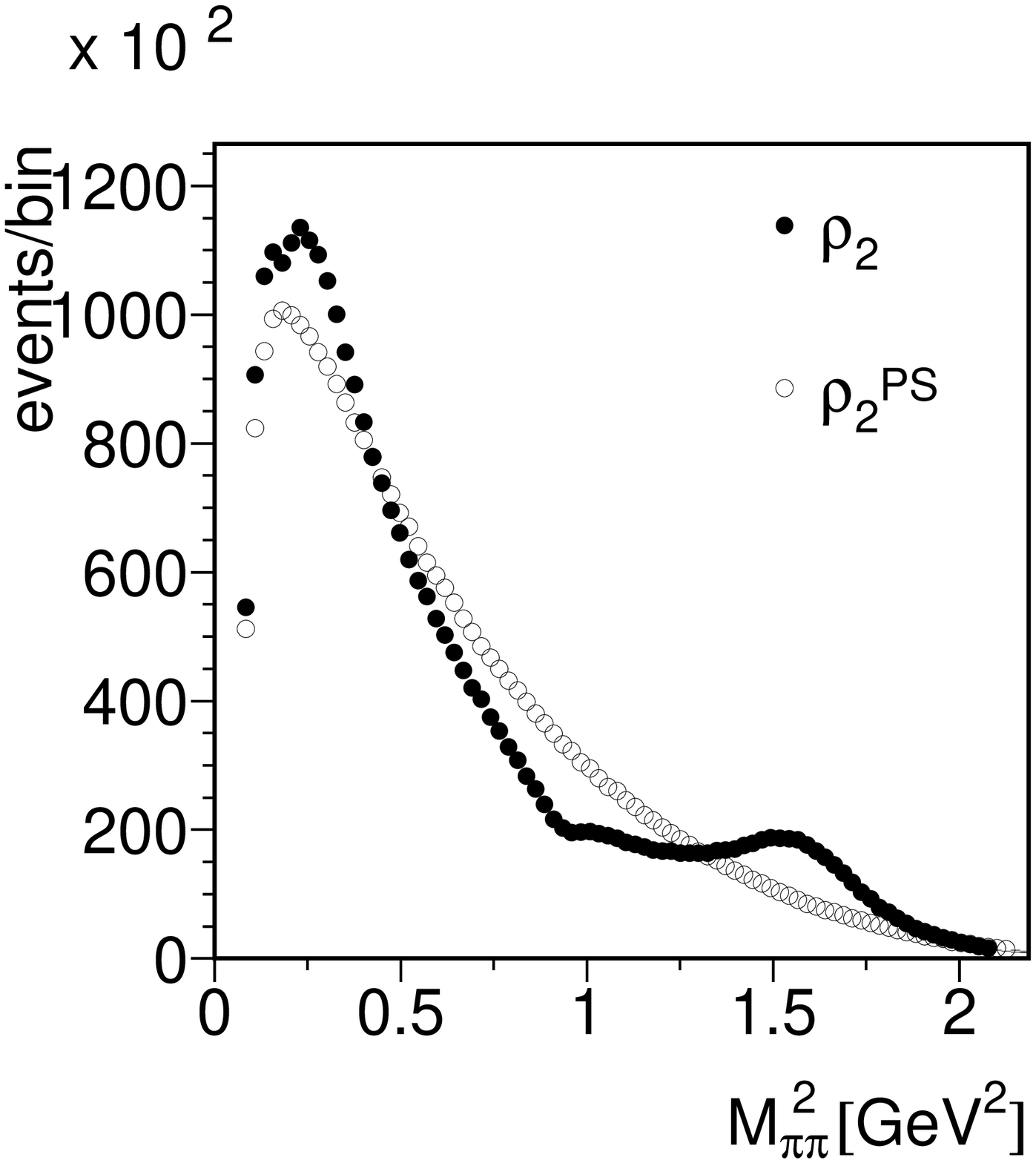}}}
  \parbox{7cm}{\center(b)\\ \mbox{\epsfysize=7cm\epsffile{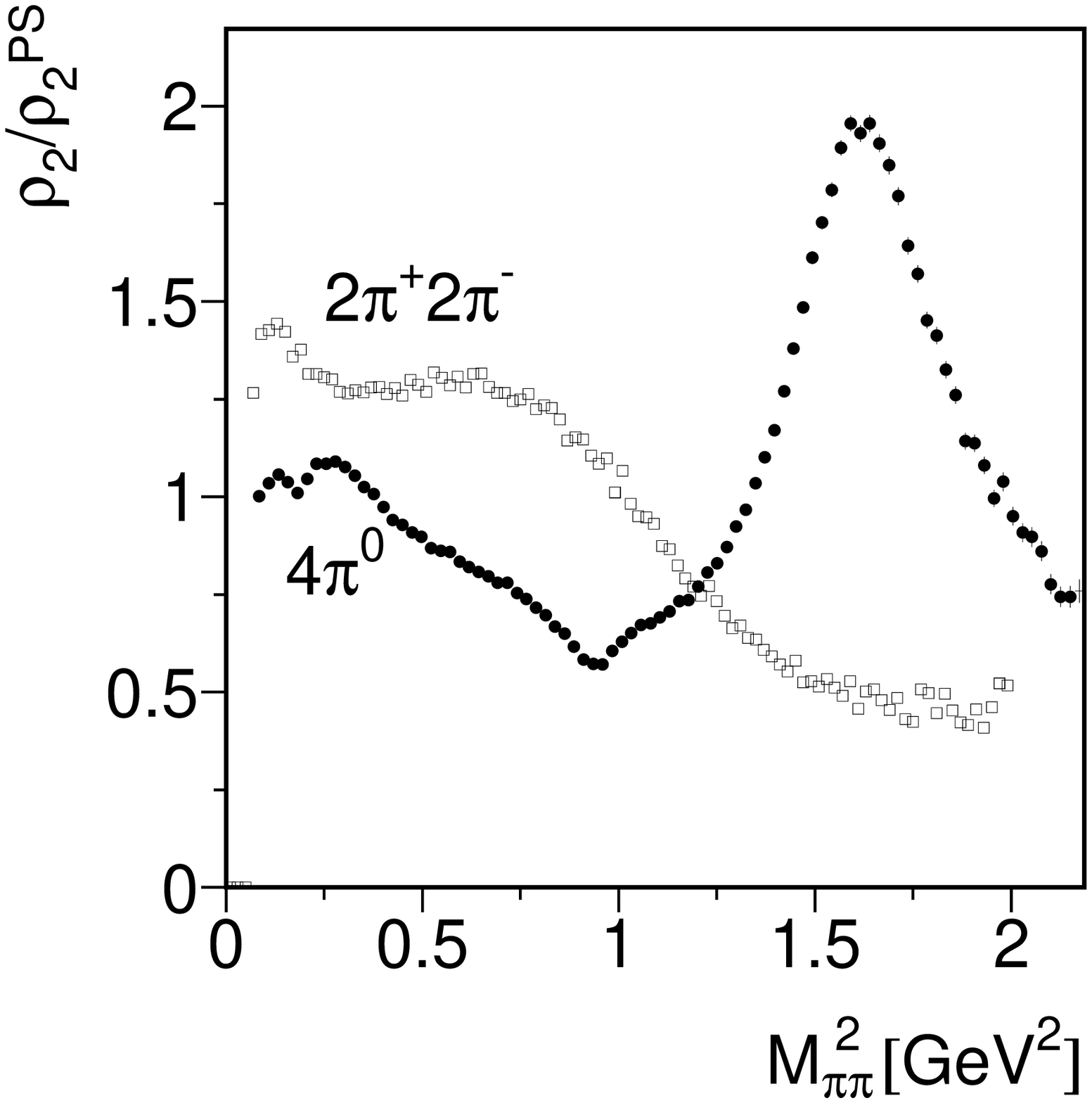}}}
}
\caption{
  Inclusive two-pion correlations 
  {\it vs.} the effective mass squared $M^2$ of the pion pair 
  in the $4\pi^0$ channel: 
  \quad
  (a) The experimental two-particle distribution $\rho_2(M)$, Eq.(\ref{rho2}),
      in comparison with the phase space distribution $\rho_{2}^{PS}(M)$.
      The simulated spectrum is normalized to the number of measured events.
  \quad
  (b) The ratio $\rho_2(M)/\rho_{2}^{PS}(M)$. 
      The $2\pi^+2\pi^-$ data are from \protect\cite{CPLEAR97}.   
} 
\label{Figrho2}
\end{figure}

The data sample of 459803 events was used to calculate 
the two-particle distributions $\rho_2^{00}(M)$ 
defined by Eq.(\ref{rho2}) for $\pi^0$ 
pairs\footnote{Here and below all distributions for $4\pi^0$ events 
contain multiple entries per event corresponding to all possible
two--particle combinations.}, see Fig.~\ref{Figrho2}a. 
  The corresponding two-particle density from the phase space simulation is
called $\rho_2^{PS}(M)$.  
The simulation was produced with the GEANT software \cite{GEANT},  
and subjected to the same cuts and selections as the real data.  
It comprises 956511 events. 
Figure~\ref{Figrho2}b shows the ratio $\rho_2^{00}(M)/\rho_2^{PS}(M)$,  
for which the kinematical correlations discussed in section~\ref{CF} cancel.
An interesting feature of this result is that there is no familiar BE peak 
at small invariant mass of the pion pair, contrary to the case 
of $\ppb\to 2\pi^+2\pi^-$ \cite{CPLEAR97}, which is shown for comparison 
in Fig.~\ref{Figrho2}b.   
The peak at $M=1.6\;$\GeV\ is due to $f_2\to\pi^0\pi^0$. 

\begin{figure}[htbp]
\centering
\mbox{\hspace*{-10mm}
  \parbox{7cm}{\center(a)\\ \mbox{\epsfysize=7cm\epsffile{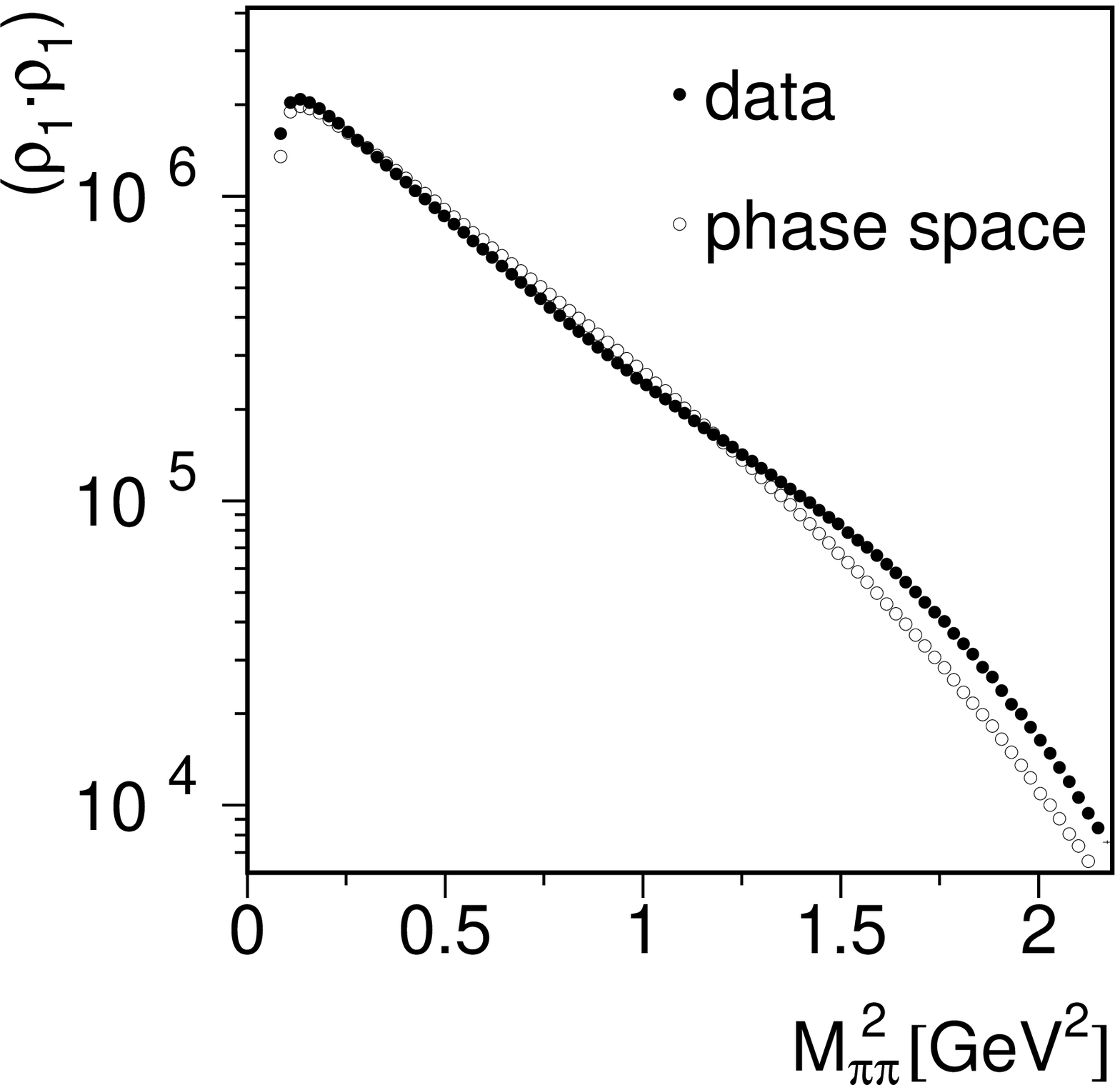}}}
  \parbox{7cm}{\center(b)\\ \mbox{\epsfysize=7cm\epsffile{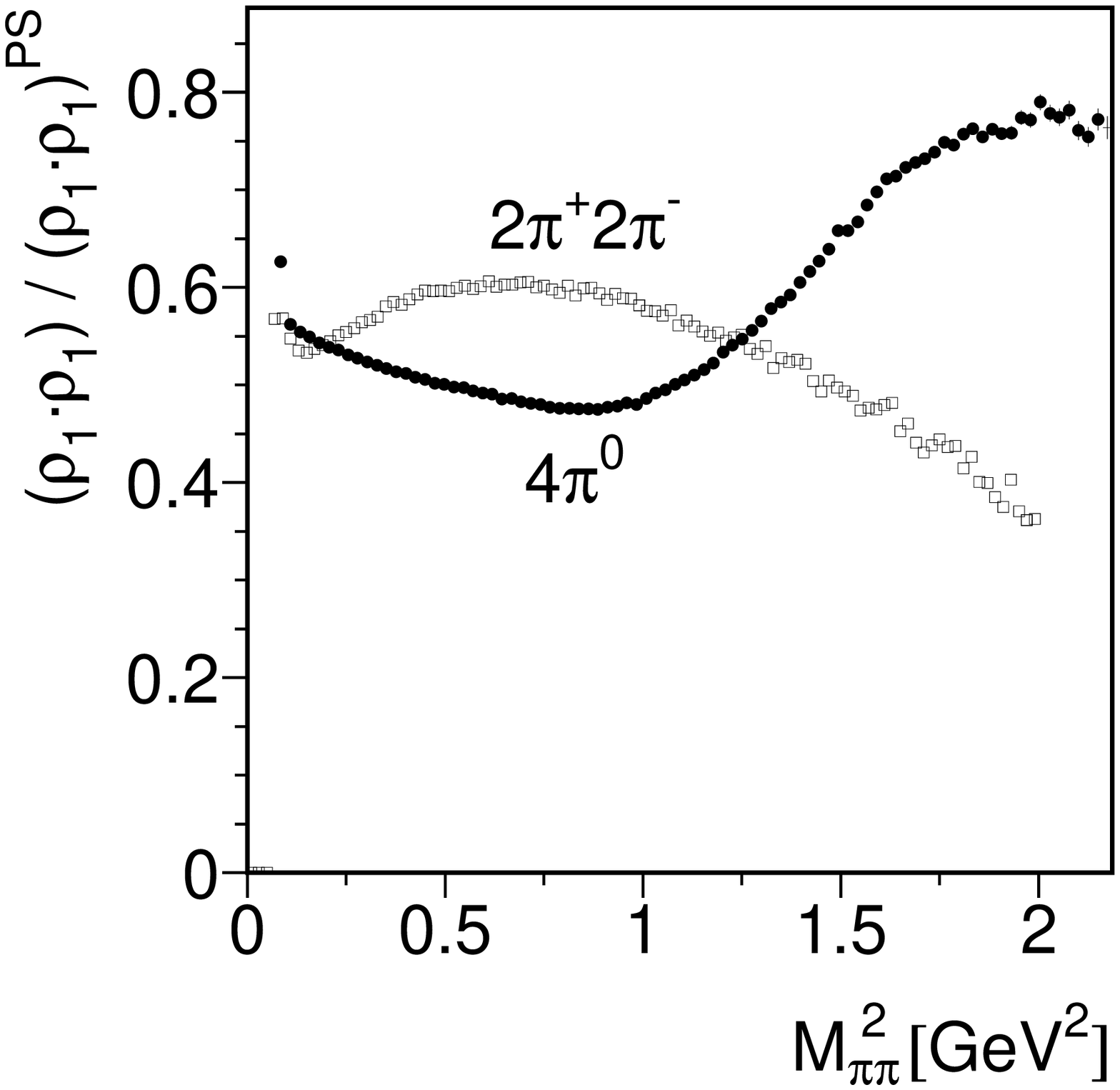}}}
}
\caption{
  a) The experimental distribution $(\rho_1\cdot\rho_1)(M)$
     ($\protect\bullet$)   
     and the phase space distribution $(\rho_1\cdot\rho_1)^{PS}(M)$
     ($\protect\circ$)
     obtained by mixing events.
     \quad 
  b) The ratio \mbox{$(\rho_1\cdot\rho_1)(M)/(\rho_1\cdot\rho_1)^{PS}(M)$}.  
     The simulated spectrum is normalized to the number of measured events.
     The $2\pi^+2\pi^-$ data are from \protect\cite{CPLEAR97}.   
}
\label{Figrho1rho1}
\end{figure}

\begin{figure}[htbp]
\centering
\mbox{\hspace*{-10mm}
  \parbox{7cm}{\center(a)\\ \mbox{\epsfysize=7cm\epsffile{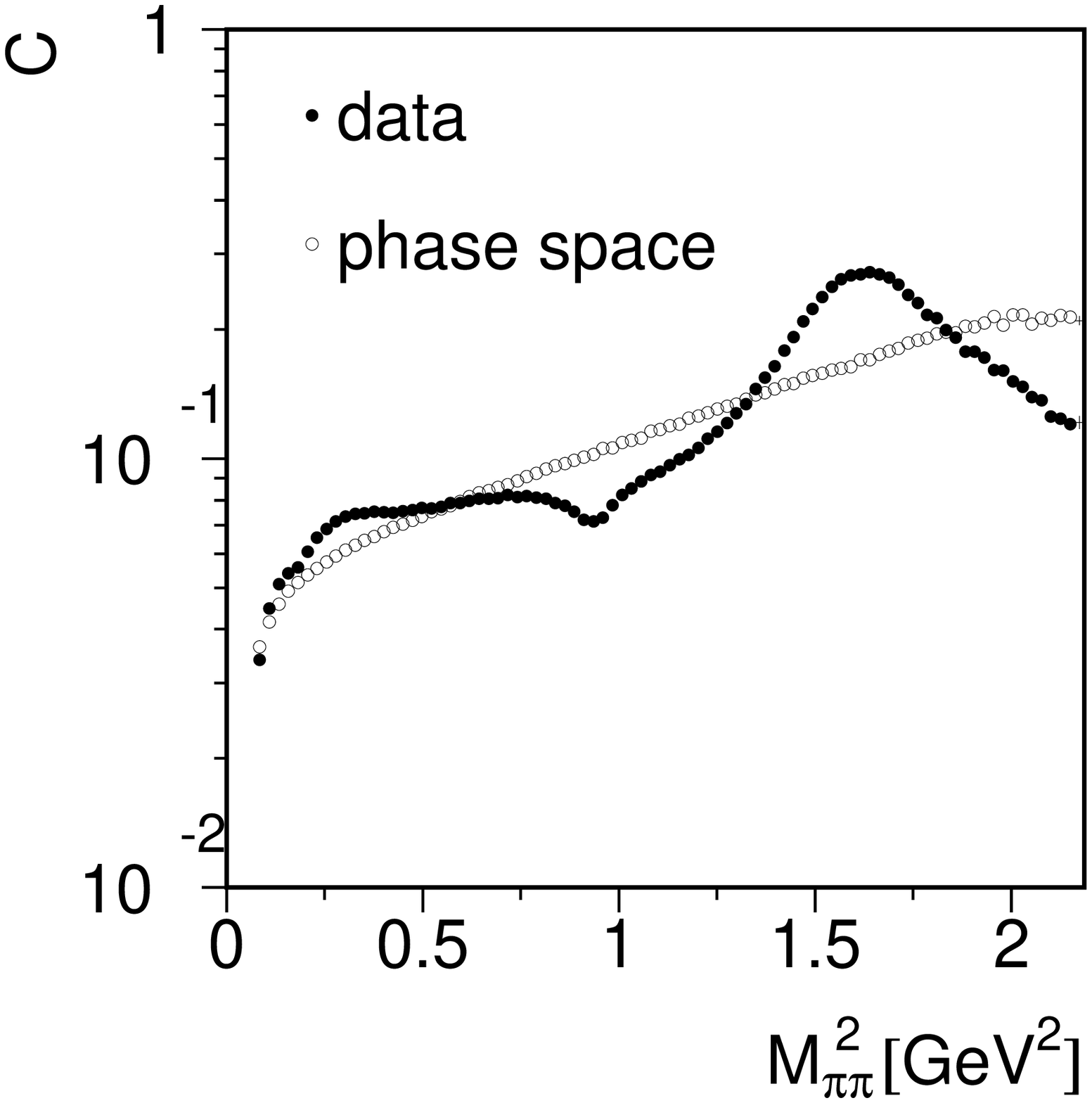}}}
  \parbox{7cm}{\center(b)\\ \mbox{\epsfysize=7cm\epsffile{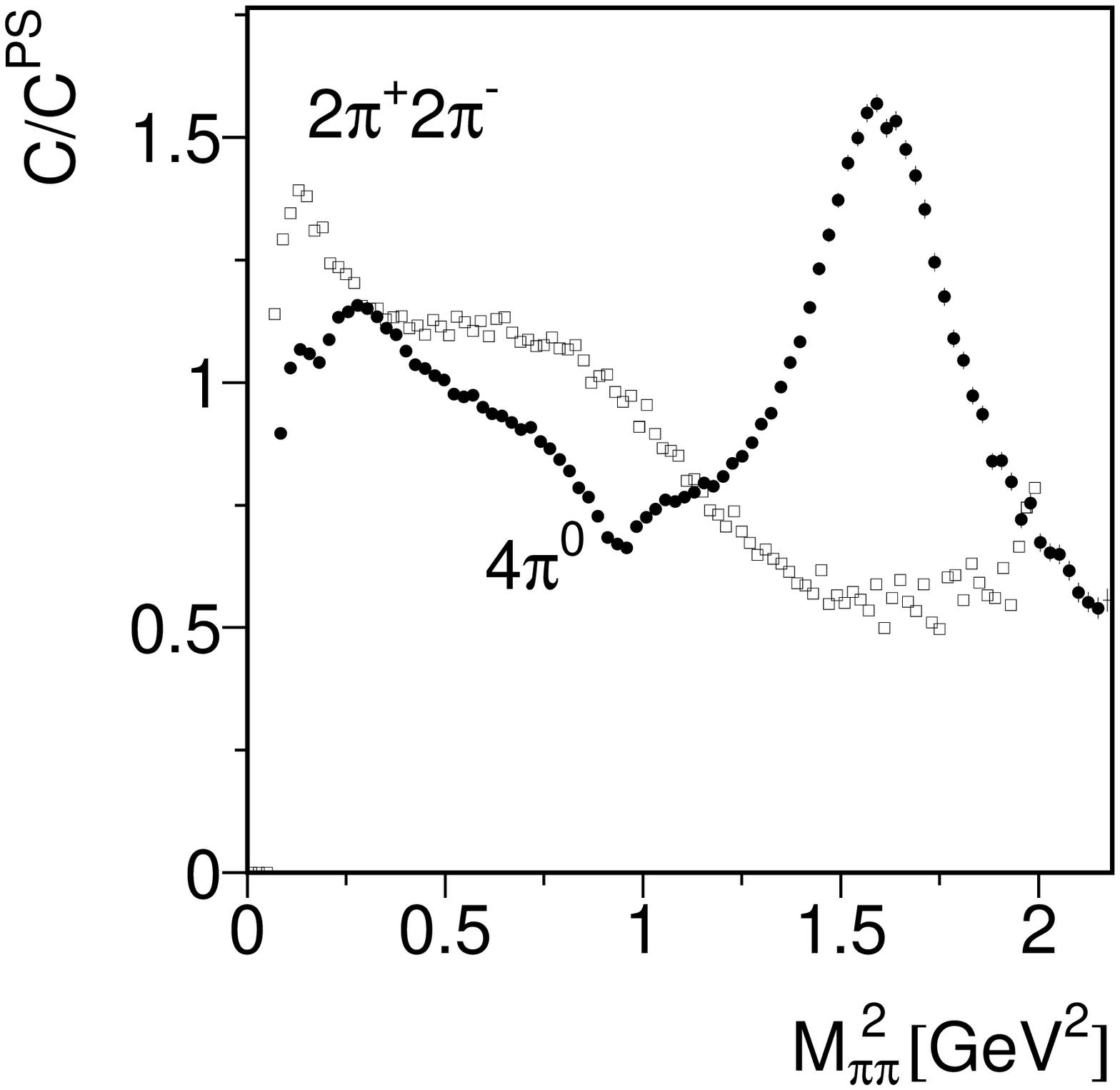}}}
}
\caption{ \label{c2exp}
(a) The experimental correlation function $C(M)$, Eq.(\ref{CM}), {\it vs.}
  the square of the effective mass $M^2$ of two pions in the $4\pi^0$ channel  
  ($\protect\bullet$)   
  in comparison with the corresponding phase space correlation function 
  ($\protect\circ$).
  \quad 
(b) The experimental correlation function normalized to the phase space
  distribution, $C(M)/C^{PS}(M)$, Eq.(\protect\ref{CexpCps}). 
  The $2\pi^+2\pi^-$ data are from \protect\cite{CPLEAR97}.   
} 
\end{figure}

  To study the correlation function $C(M)$ in Eq.(\ref{CM}), 
the two-particle distribution for uncorrelated pion pairs has been 
calculated using the same event-mixing method as in \cite{CPLEAR97}. 
  The experimental distribution $(\rho_1\cdot\rho_1)(M)$
and the simulated distribution $(\rho_1\cdot\rho_1)^{PS}(M)$
are plotted in figure~\ref{Figrho1rho1}, along with their ratio.
  Figure~\ref{c2exp} shows the correlation functions $C(M)$ for
the $\pi^0\pi^0$ pairs in comparison with the phase space distribution.
  In order to remove the trivial $M$-dependence which arises 
from the energy-momentum conservation, equations (\ref{CM}-\ref{rho11}), 
and the influence of the 
experimental cuts, the following double ratio has been calculated:
\BA
  \frac{C(M)}{C^{PS}(M)} =
  \frac{\rho_2(M)}{(\rho_1\cdot\rho_1)(M)} :
  \frac{\rho_2^{PS}(M)}{(\rho_1\cdot\rho_1)^{PS}(M)}   \quad .
  \LL{CexpCps}
\EA
The result is shown in Fig.\ref{c2exp}b. 
Clearly the inclusive $\pi^0$ pair correlation function has a dip at small 
$M$. This is unexpected in the conventional HBT picture.
It also contrasts to our findings for charged pion correlations,
where the same correlation function shows a weak enhancement at
small $M$ \cite{CPLEAR97}, see Fig.\ref{c2exp}b.   
  As discussed in section~\ref{DISC},  
the structures seen in Fig.\ref{c2exp} correspond to the resonances 
$f_2(1270)$, $a_2(1660)$, and $\pi_2(1670)$,

\section{Differential two-pion correlations}
\label{METH4pi0}

   So far we have presented the inclusive correlation function
$C(M)/C^{PS}(M)$ where all kinematical variables except the 
invariant mass $M$ of one pion pair have been integrated out.
To investigate possible correlation signals in more detail we turn to 
differential densities. Our approach follows the method used
previously for the systems of charged pions \cite{CPLEAR97,CPLEAR98}. 

  The reaction $\ppb \to 4\pi^0$ at rest in the CB experiment
proceeds from the $S$-wave state $J^{PC}=0^{-+}$ and the 
$P$-wave atomic states $J^{PC}=0^{++},1^{++},2^{++}$.   
The corresponding pion distribution for the final state configuration
${\{\MBF{p}_i\}, \ i=1,2,3,4}$, has the form
\begin{eqnarray}
  d\sigma (\{\MBF{p}_i\}) & \sim &   
   \left(w_S |T(\MBF{k},\{\MBF{p}_i\})|^2 + 
         w_P |\nabla_k T(\MBF{k},\{\MBF{p}_i\})|^2 \right)_{\MBF{k}\to 0} \;
   d\Phi_4(p,p_1,p_2,p_3,p_4) 
   \NN \\
   & = & 
   \overline{|T(\MBF{k},\{\MBF{p}_i\})|^2_{\MBF{k}\to 0}} \; 
   d\Phi_4(p,p_1,p_2,p_3,p_4) 
   \quad .
\LL{dW}
\end{eqnarray}
Here $T(\MBF{k},\{\MBF{p}_i\})$ is the amplitude of the $\ppb$ 
annihilation from the initial $\ppb$ state with relative
momentum $\MBF{k}$,
$d\Phi_4(p,p_1,p_2,p_3,p_4)$ is the $4$-body relativistic phase space, 
and the limit $\MBF{k}\to 0$ implies the incoherent sum\footnote{%
The notation $\overline{|T(\MBF{k},\{\MBF{p}_i\})|^2_{\MBF{k}\to 0}}$  
implies a sum over initial spin states 
$J^{PC}=0^{-+},0^{++},1^{++},2^{++}$, and all quantum numbers specifying the 
initial spin and the total angular momentum are suppressed.} 
of the $S$ and $P$-wave states with the corresponding weights $w_S$ and $w_P$.
The four-vectors of the pions are $p_i=(E_i,\MBF{p}_i)$, and 
$p=(2m_p,0)$ is the total four-momentum for $\ppb$ annihilation at rest,
$m_p$ being the proton mass.

\begin{figure}[htbp]
\centering
\mbox{\epsfxsize=4cm \epsffile{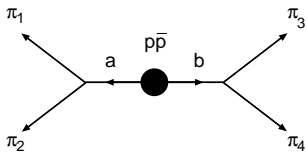}} 
\caption{The two-pion subsystems $a=(\pi_1+\pi_2)$ and $b=(\pi_3+\pi_4)$ 
in the four pion final state.} 
\label{Scheme4pi}
\end{figure}

  We introduce the two-pion subsystems $a$ and $b$ with four-momenta
$p_a=(p_1+p_2)$ and $p_b=(p_3+p_4)$ and invariant masses $M_a$ and $M_b$,
see figure~\ref{Scheme4pi}. 
Given the invariant masses $M_a$ and $M_b$, the double 
differential cross section is defined by integrating over the angles
specifying the relative orientation of the momenta of the pions within the
subsystems $a$ and $b$ and the relative orientation of the subsystems $a$ and
$b$ (the corresponding solid angles
are $d\Omega_{12}$, $d\Omega_{34}$, $d\Omega_{ab}$).  
For the reaction $\ppb \to 4\pi^0$ we have 
\BA
   \frac{\displaystyle d \sigma}{\displaystyle dM_a^2 dM_b^2}
   & \sim &   
   W(\sqrt{s},M_a,M_b) 
   \int \overline{|T(\MBF{k},\{\MBF{p}_i\})|^2_{\MBF{k}\to 0}}
   d\Omega_{ab} d\Omega_{12} d\Omega_{34} \quad .  
\LL{WTI}
\EA
Here the factor $W(M,M_a,M_b)$ is given by 
\BA
  W(M,M_a,M_b) & = &
   \frac{P_{ab}}{M}    
   \sqrt{\left(1-\frac{4\mu^2}{M_a^2}\right)
         \left(1-\frac{4\mu^2}{M_b^2}\right)}  \quad , \LL{Wab}
\\
  P_{ab} & = & P(M,M_a,M_b) \quad , \LL{Pab}
\\ 
  P(M,M_a,M_b) & = & 
       \frac{\sqrt{(M^2-(M_a+M_b)^2)(M^2-(M_a-M_b)^2)}}{2M}
\EA
where $P(M,M_a,M_b)$ is the relative momentum of two particles with 
masses $M_a$ and $M_b$ and the total invariant mass $M$, 
in our case $M=2\Mp$.  
Removing the phase space factor $W(M,M_a,M_b)$, 
we define the double--differential density:
\begin{eqnarray} 
  \RHO(M_a,M_b) & = &
  \frac{1}{W(M,M_a,M_b)} \;  
  \frac{d\sigma}{\sigma \cdot d M_a^2 d M_b^2}  \label{rhoW} \NN \\
  & \sim &
  \int \overline{|T(\MBF{k},\{\MBF{p}_i\})|^2_{\MBF{k}\to 0}} \;  
  d\Omega_{ab} d\Omega_{12} d\Omega_{34} 
   \quad 
\label{rhosig} 
\end{eqnarray}
where $\sigma$ is the total cross section. 

An advantage of using the double differential density $\RHO(M_a,M_b)$
is that it does not contain {\it kinematical} dependences on the
invariant masses of the two-pion pairs $a$ and $b$.  This means that
for a constant $T$ matrix the density
$\RHO(M_a,M_b)$ does not depend on its argument $M_a$ and $M_b$,
contrary to $\rho_2(M_a) = d\sigma/d M_a$ of Eq.(\ref{rho2}) 
which unavoidably involves the phase--space dependence.

\begin{figure}[htbp]
\centering
\mbox{\hspace*{-10mm}
  \parbox{7cm}{\center(a)\\ \mbox{\epsfysize=7cm\epsffile{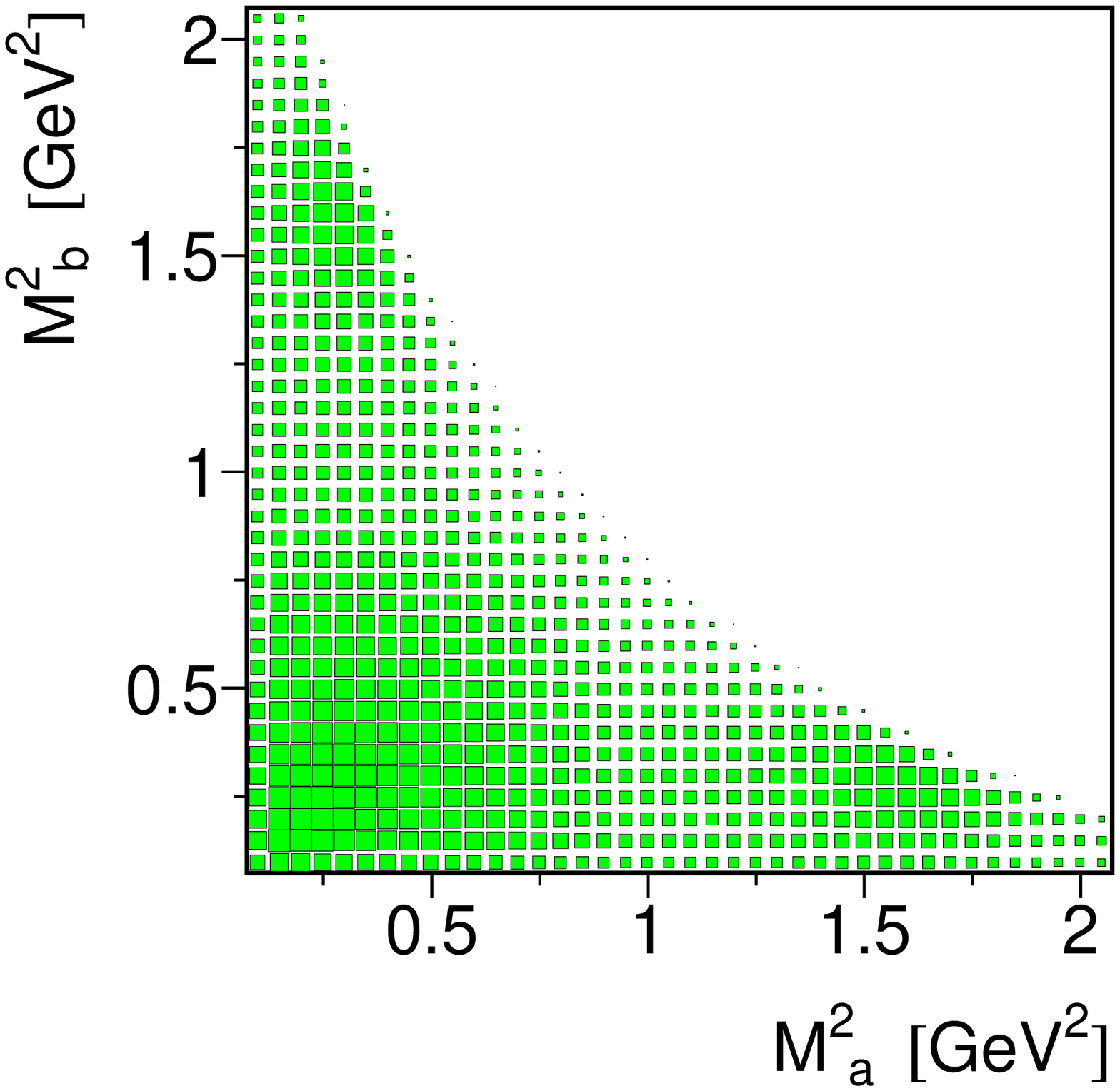}}}
  \parbox{7cm}{\center(b)\\ \mbox{\epsfysize=7cm\epsffile{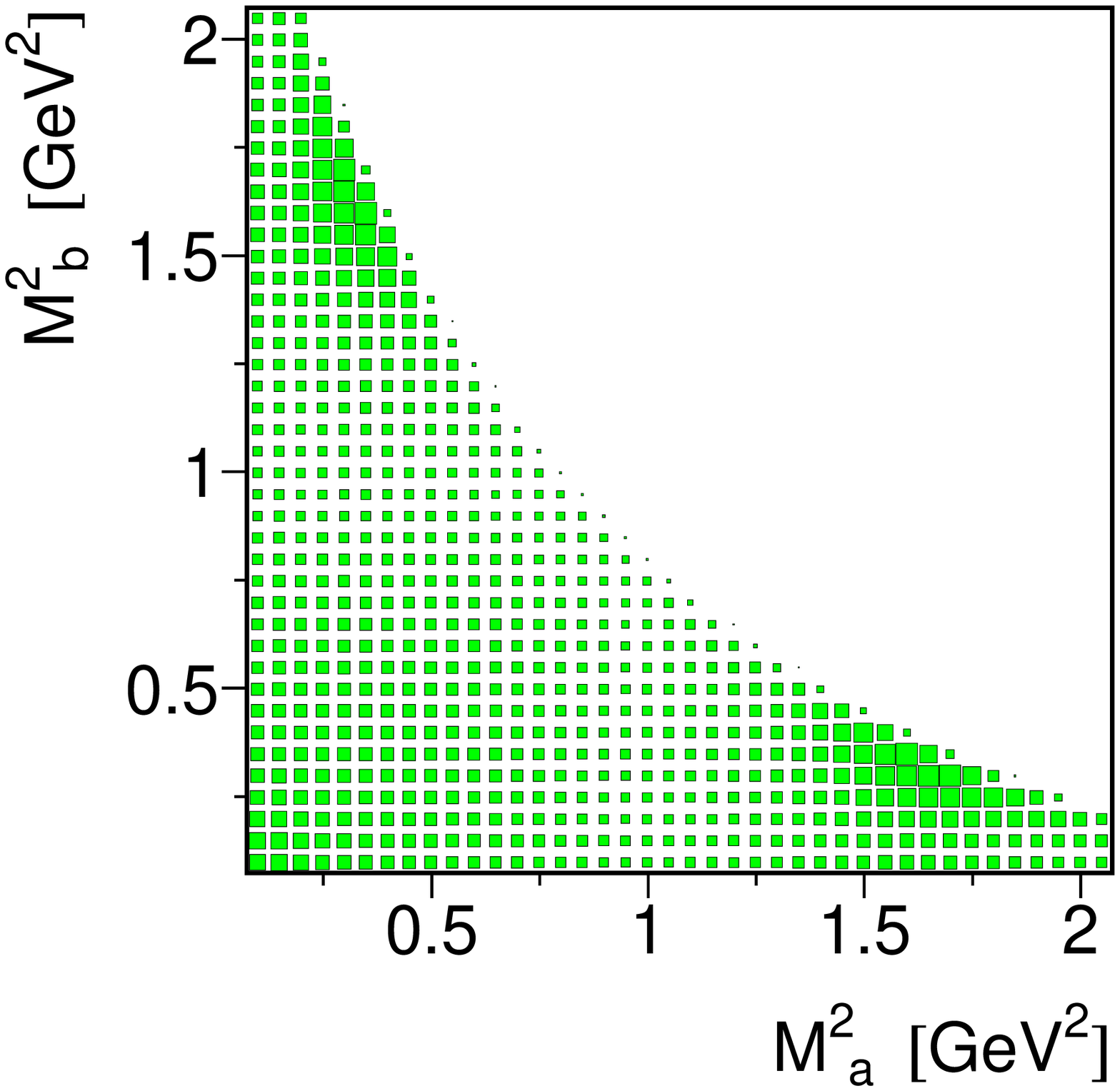}}}
}
\mbox{\hspace*{-10mm}
  \parbox{7cm}{\center(a)\\ \mbox{\epsfysize=7cm\epsffile{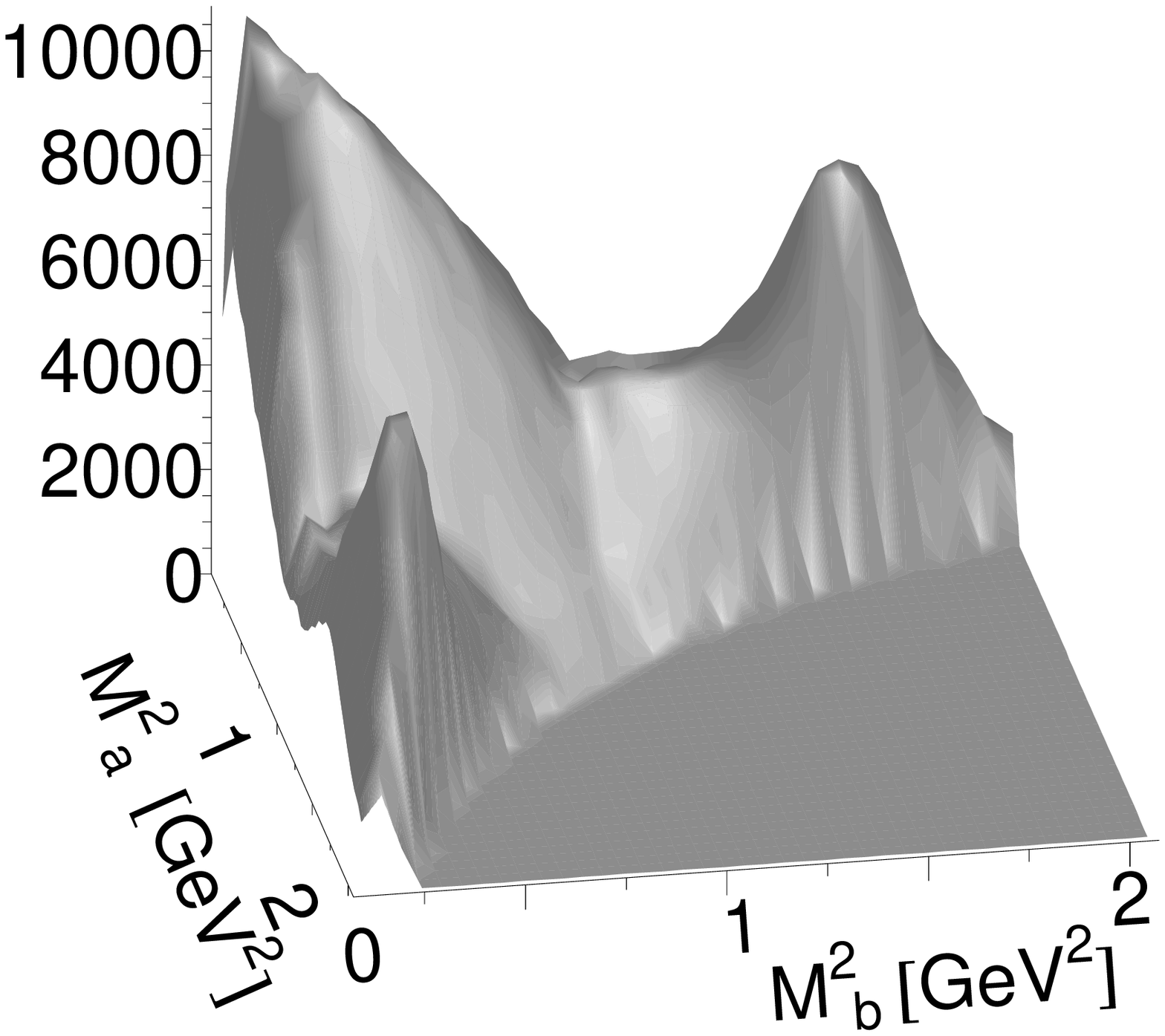}}}
  \parbox{7cm}{\center(b)\\ \mbox{\epsfysize=7cm\epsffile{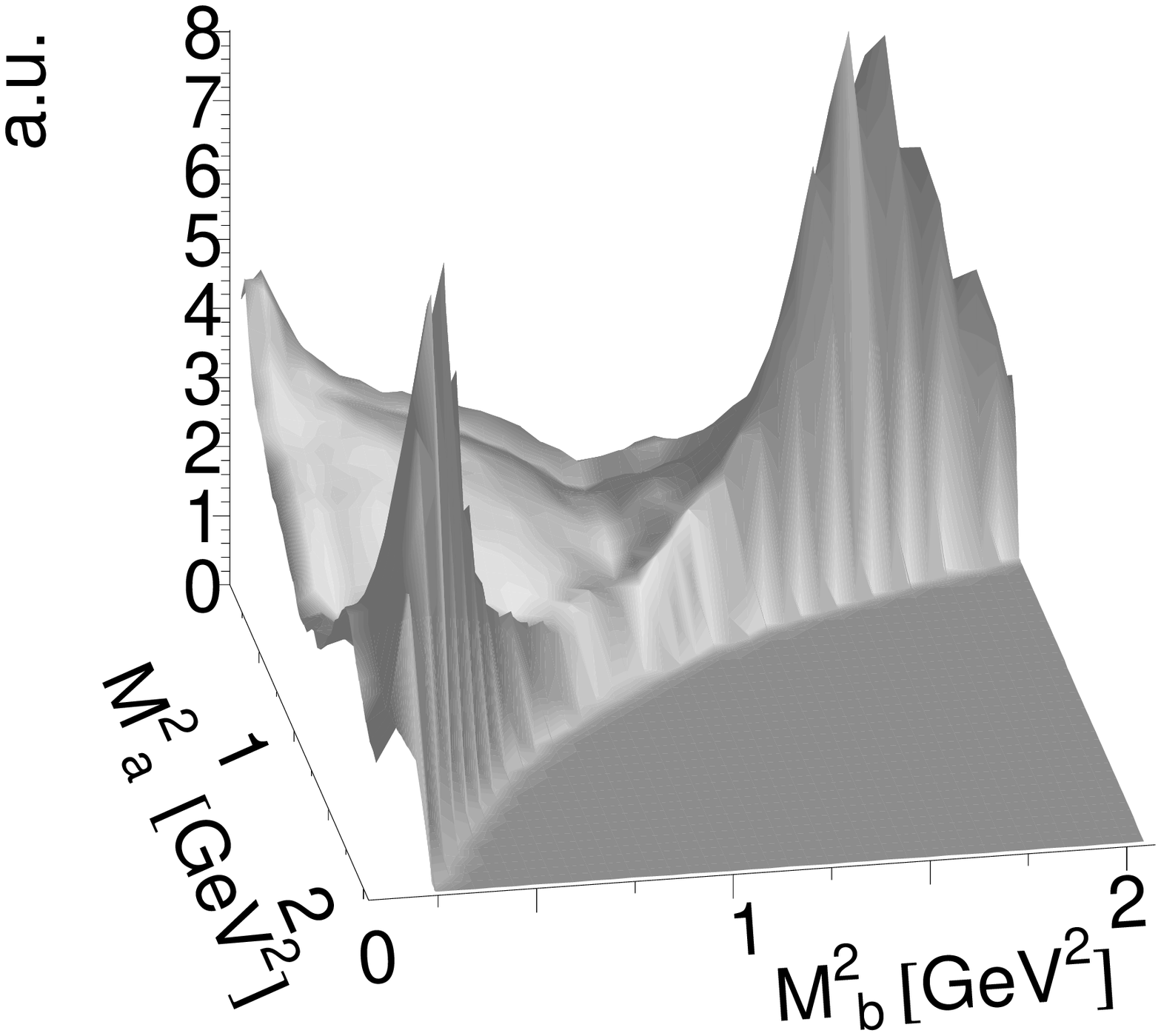}}}
}
\caption{ \label{sr40}
   The double-differential distributions 
   {\it vs.} the invariant masses of two $\pi^0$ pairs 
   (6 entries per one physical event): 
  \quad
   (a) the double-differential cross section
       $d\sigma/d\Mooa^2 d\Moob^2$;    
   \quad
   (b) the double-differential density 
      $\RHO(\Mooa,\Moob)$, Eq.~(\protect\ref{rhosig}).
   The box plots (top) and the surface plots (bottom)
   represent the same data. 
   Both the cross section and the density plots are 
   not corrected for acceptance. 
}
\end{figure}

\begin{figure}[htbp]
\centering
\mbox{\hspace*{-10mm}
  \parbox{7cm}{\center(a)\\ \mbox{\epsfysize=7cm\epsffile{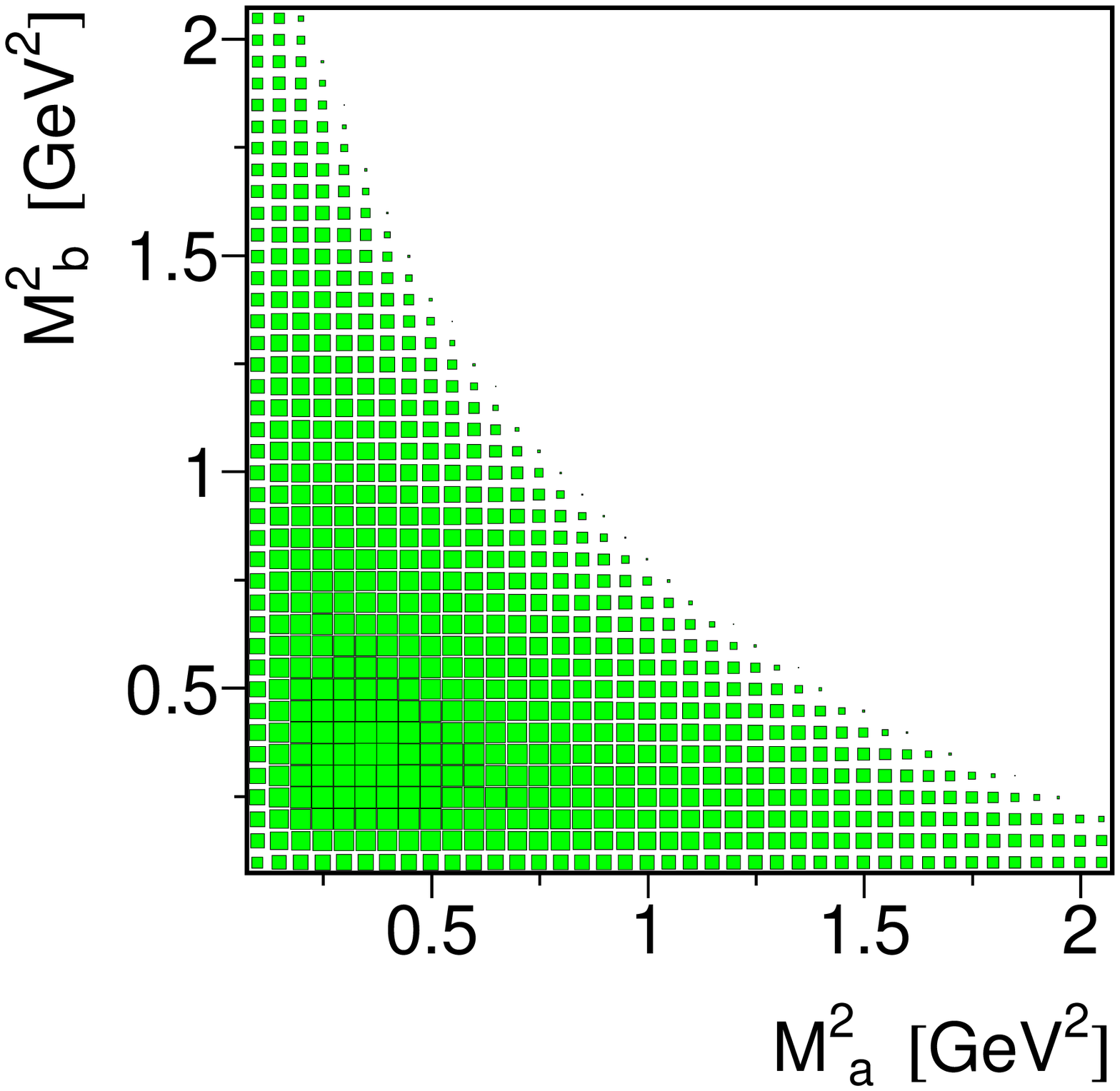}}}
  \parbox{7cm}{\center(b)\\ \mbox{\epsfysize=7cm\epsffile{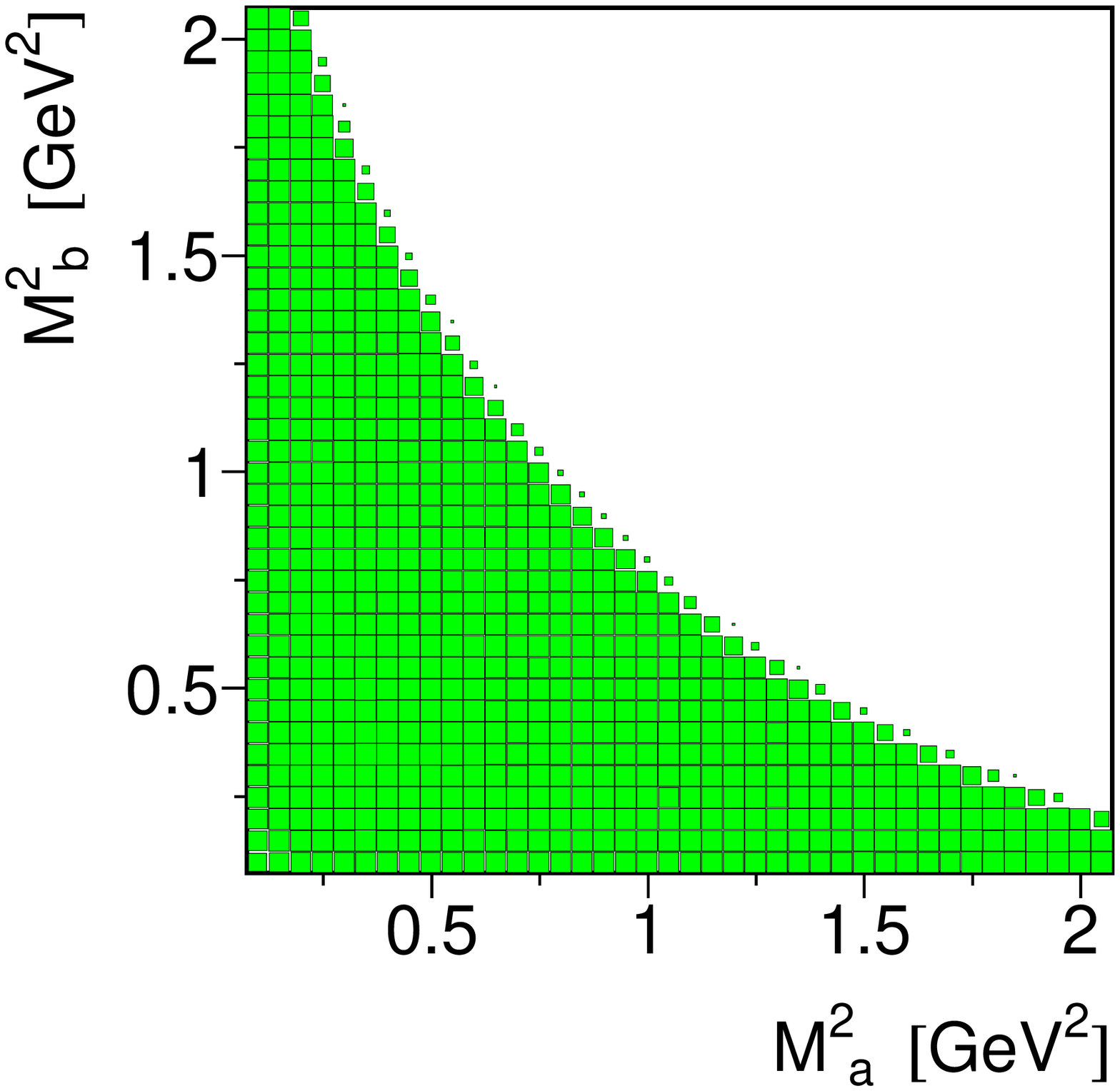}}}
}
\mbox{\hspace*{-10mm}
  \parbox{7cm}{\center(a)\\ \mbox{\epsfysize=7cm\epsffile{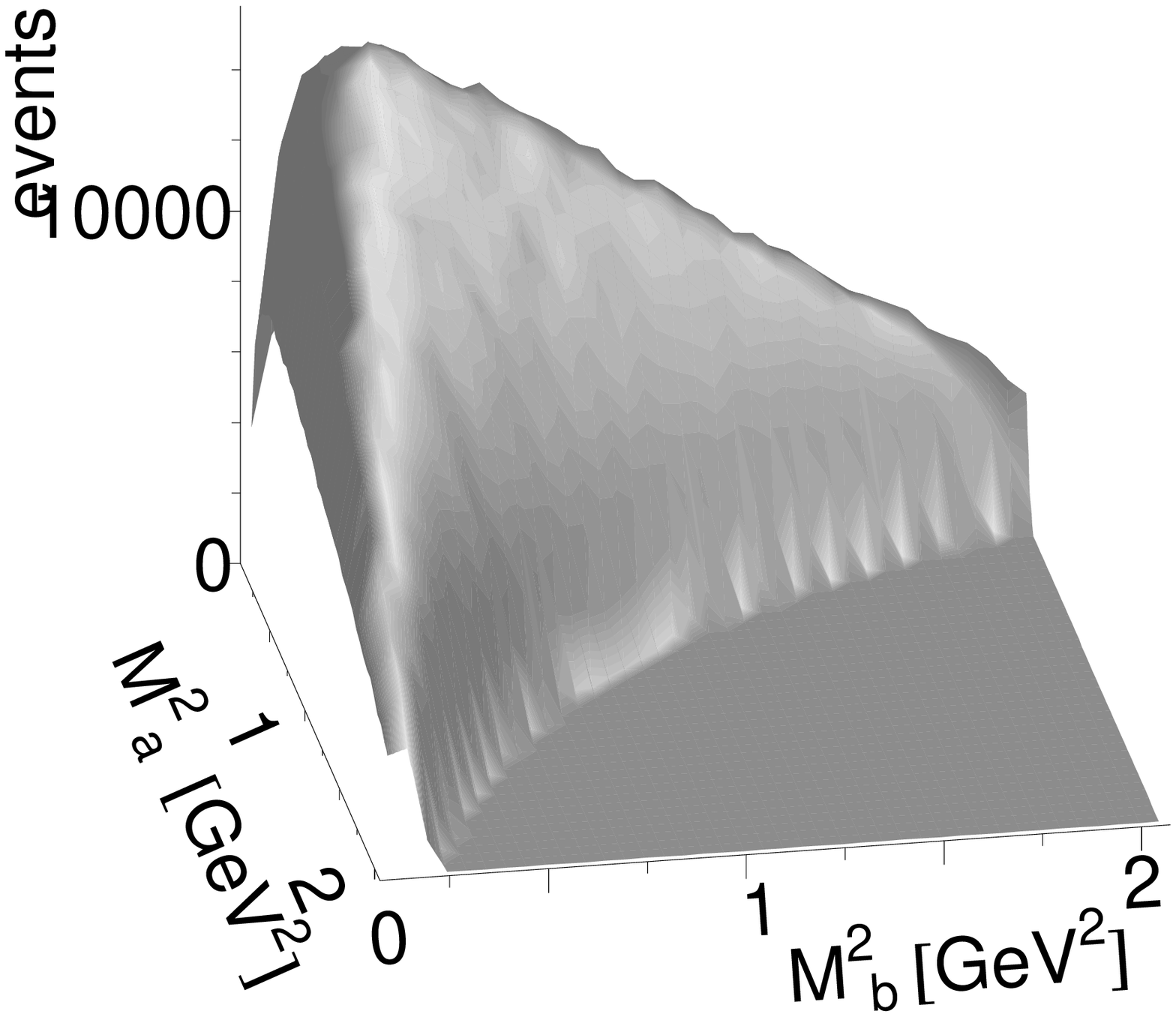}}}
  \parbox{7cm}{\center(b)\\ \mbox{\epsfysize=7cm\epsffile{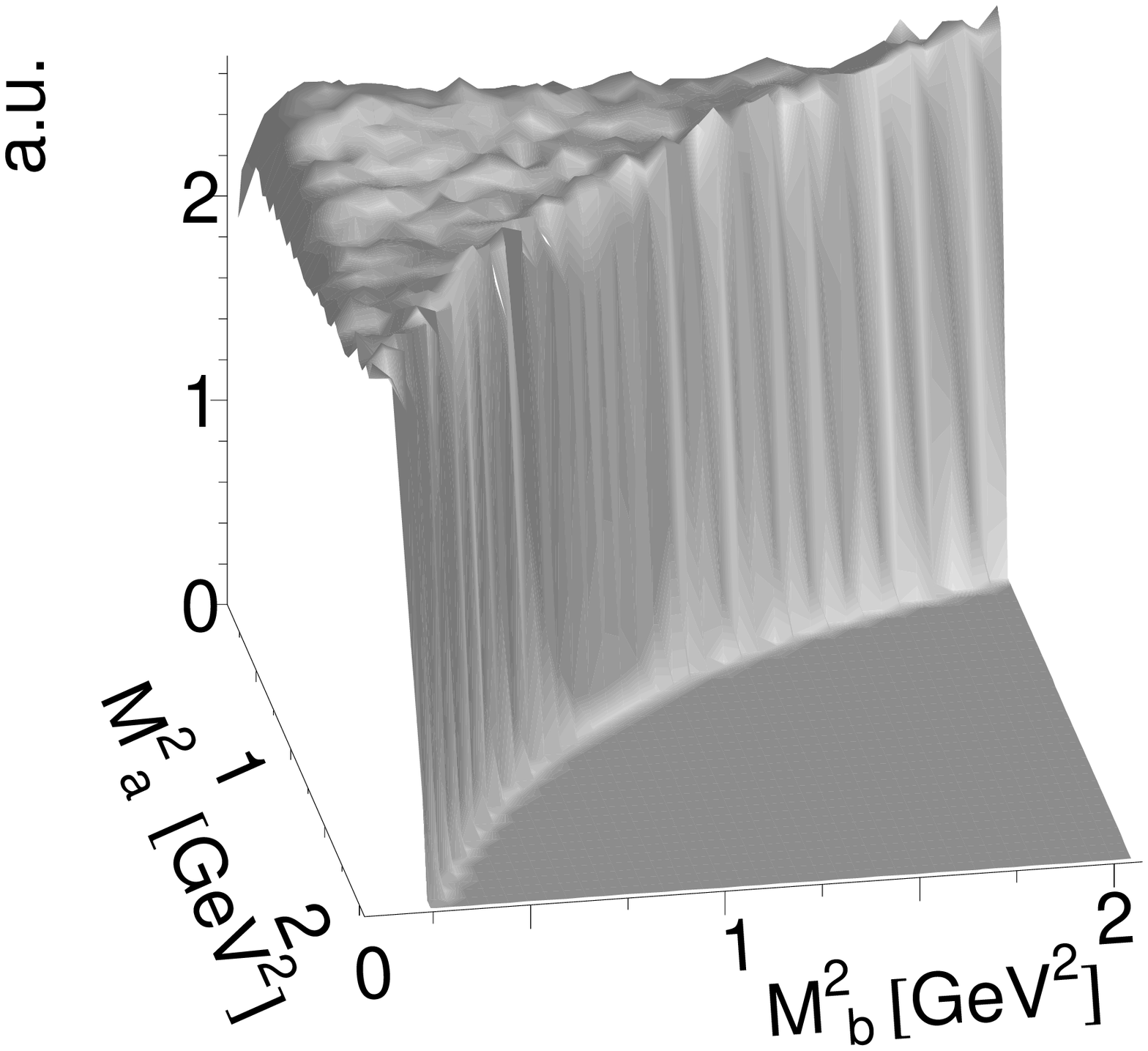}}}
}
\caption{ \label{sm40}
  The acceptance simulation of the detector,
  calculated for events distributed uniformly in phase space: 
  \quad
  (a) the double-differential distribution
      $d\sigma^{\mathrm{PS}}/d\Mooa^2 d\Moob^2$; 
  \quad
  (b) the double-differential density $\RHO^{\mathrm{PS}}(\Mooa,\Moob)$.
  The box plots (top) and the surface plots (bottom)
  represent the same data.
}
\end{figure}

\begin{figure}[htbp]
\centering 
\mbox{\hspace*{-10mm}
  \mbox{\epsfysize=7cm\epsffile{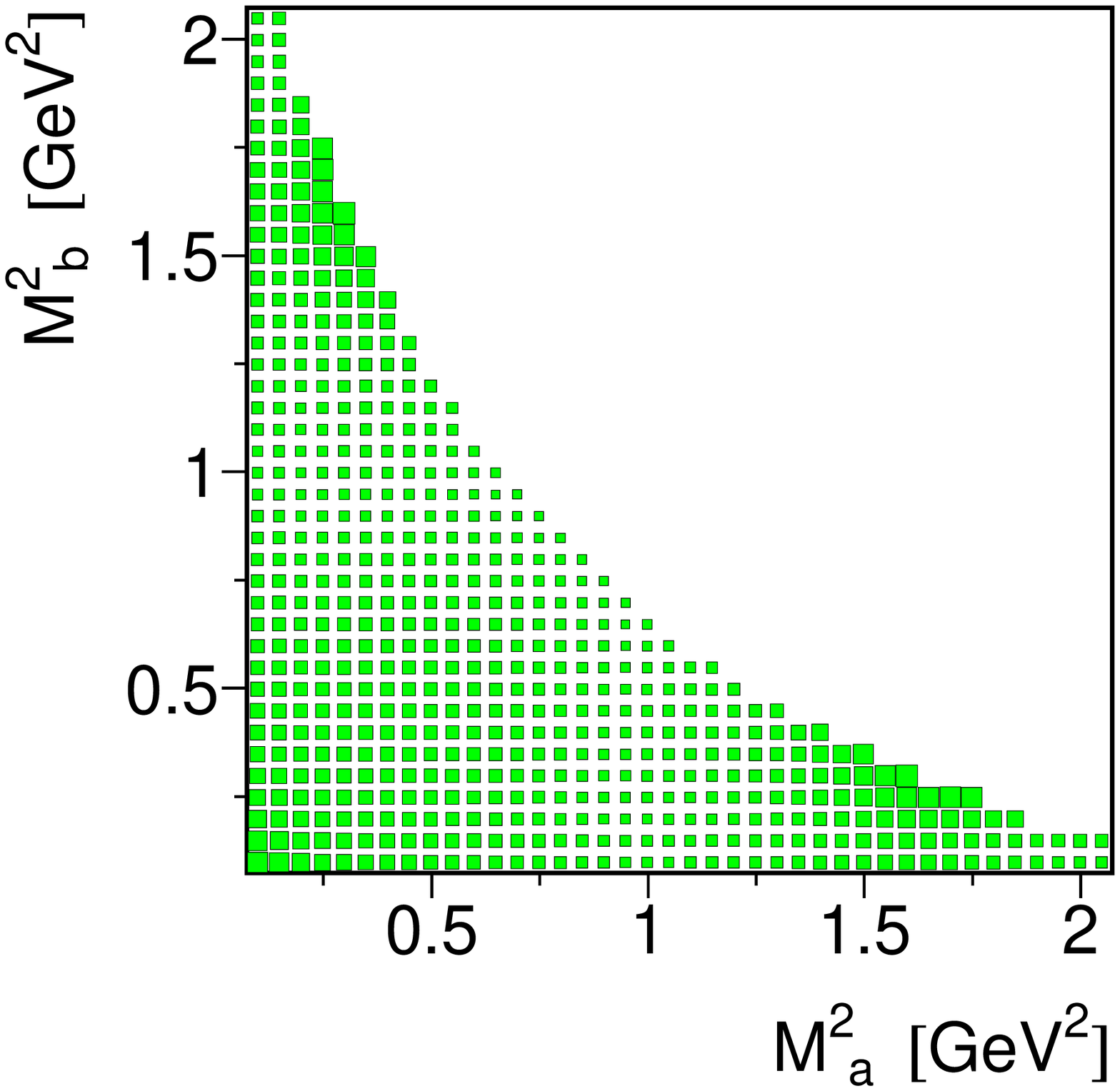}} 
  \mbox{\epsfysize=7cm\epsffile{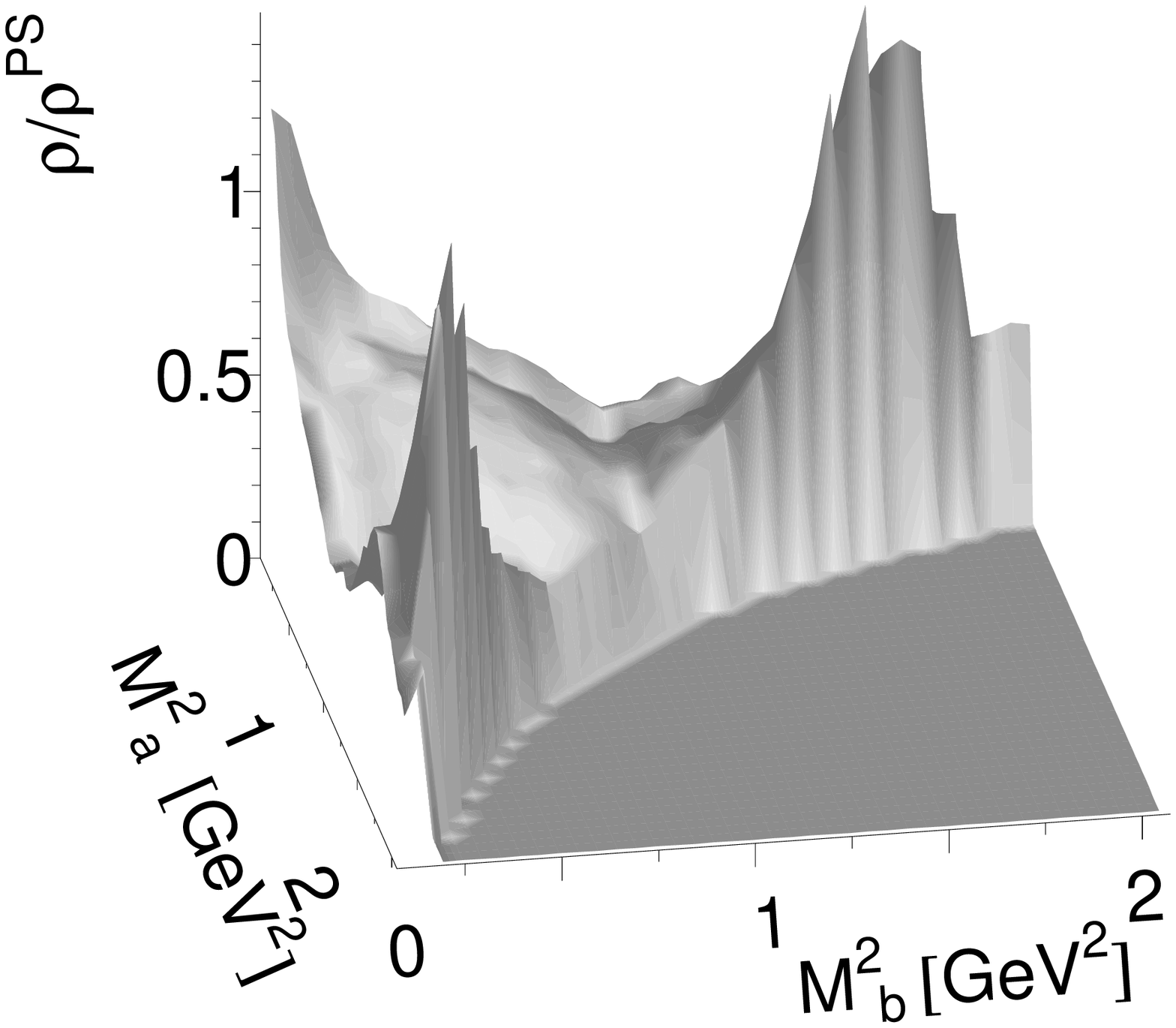}} 
}
\caption{ \label{rg40}
  The experimental double--differential density
  $\RHO(\Mooa,\Moob) / \RHO^{\mathrm{PS}}(\Mooa,\Moob)$
  corrected for acceptance.
  Both plots represent the same data.
}
\end{figure}

The double-differential cross-section $d\sigma/dM_{a}^2 dM_{b}^2$
corresponding to the raw data is shown in figure~\ref{sr40}a, and the
corresponding double differential density $\RHO(\Mooa,\Moob)$ in
figure~\ref{sr40}b. 
These raw data must be corrected for the acceptance of the detector, 
which is shown in figure~\ref{sm40}.    
The differential density $\RHO^{\mathrm{PS}}(\Mooa,\Moob)$ is calculated
using the four--pion phase space with the detector acceptance taken into
account. 
The double differential density corrected for the detector acceptance 
$\RHO(\Mooa,\Moob) / \RHO^{\mathrm{PS}}(\Mooa,\Moob)$ is plotted in
figure~\ref{rg40}.  
Note that the double differential view shows a weak enhancement at low
values of $M^2$ which is absent in the inclusive projection of
figure~\ref{c2exp}.

\begin{figure}[htbp]
\centering
\mbox{\hspace*{-10mm}
  \mbox{\epsfysize=14cm \epsffile{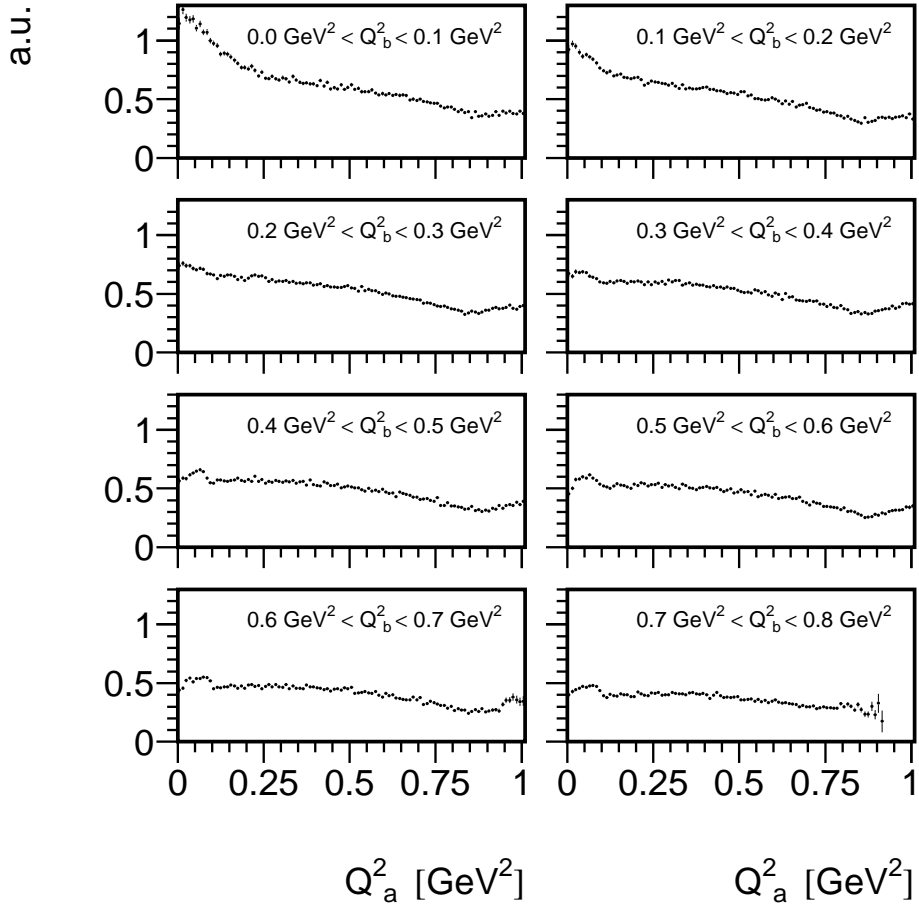}}
}
\caption{ \label{ppr40}
  The acceptance--corrected 
  projections $\RHO_i(\Mooa)$ of the differential density
  $\RHO(\Mooa,\Moob) / \RHO^{\mathrm{PS}}(\Mooa,\Moob)$
  for different intervals of the invariant mass
  of the other $\pi^0\pi^0$ pair.
  The relative momentum $Q$, Eq.(\ref{MQ}), is used instead
  of the invariant mass $M$ for the sake of convenience.
}
\end{figure}

For the further discussion in section~\ref{DISC} we construct the
following partial projections of the two--particle density.
The two--dimensional space $(\Mooa,\Moob)$ is divided into slices  
$M_i^2 \leq \Moob < M_{i+1}^2$ and the projections $\RHO_i(\Mooa)$
are defined by
\BA
    \RHO_i(\Mooa) & = & 
    \int_{M_i^2}^{M_{i+1}^2} 
    \frac{\RHO(\Mooa,\Moob)}{\RHO^{\mathrm{PS}}(\Mooa,\Moob)} \; d\Moob^2
\EA 
The partial projections $\RHO_i(\Mooa)$ are plotted in Fig.~\ref{ppr40}.
Contrary to the inclusive distribution figure~\ref{c2exp}b where
a dip was observed, we see a peak in the first two slices.
The slices in the mass interval $0.3\;GeV^2 \leq \Moob^2 \leq 1.2\;\GeV^2$ 
are more or less flat.  The dip seen in Fig.~\ref{c2exp}b results from 
integration over the higher mass region of $\Moob^2 \geq 1.2\;\GeV^2$. 
  In comparison to the corresponding projections for the 
annihilation into $2\pi^+2\pi^-$, 
the observed enhancement at small $Q^2_a$ in the $\pi\pi^0$ case  
is weak even for the lowest interval  $0.0<Q^2_b<0.1 \GeV^2$.

\section{Discussion}
\label{DISC}

\subsection{Resonance effects}
\label{SRes}
\newcommand{\SEEA}{\begin{picture}(16,16)(0,0) 
                   \put(0,12){\vector(2,-1){12}} 
                   \end{picture}}

  Resonances are strongly produced in the reaction
$\ppb\to 4\pi^0$, see figures~\ref{sr40} where the signals of
$f_2$ are clearly seen in the invariant mass projections. 
The total spin $S$ and the relative angular momentum $L$ of 
the $\ppb$ system annihilating into $4\pi^0$ are constrained by
the conservation of $C$--parity: $C=(-1)^{L+S}=1$.  Therefore, 
the annihilation into $4\pi^0$ can occur in the case of    
the $S$--wave annihilation ($L=S=0$) from the state $J^{PC}=0^{-+}$ only, 
and in the case of the $P$--wave annihilation ($L=S=1$) from the states 
$J^{PC}=0^{++},1^{++},2^{++}$.    
  
Following the Crystal Barrel analysis \cite{Ko98} we focus our attention 
on the most prominent mechanisms seen in the $4\pi^0$ channel.  
For the dense gaseous target at 12~bar, the fractions of the total 
$S$-wave and $P$-wave annihilations are approximately equal to each other 
\cite{CB2001NP}.  For the sake of simplicity, we focuse our attention on one 
$P$-wave channel $J^{PC}=2^{++}$, which has the largest statistical weight, 
and the $S$-wave channel $J^{PC}=0^{-+}$.     
We do not attempt to make a global fit of the data in the $4\pi^0$ channel, 
which would be far beyond the scope of this paper, but rather to explore how  
the most probable annihilation mechanisms manifest themselves  
in the observed BE correlations. 
The following mechanisms have been taken into consideration:
\BA
   \ppb(J^{PC}=0^{-+}) & \to & a_2(1660) + \pi^0  \LL{j0ma2popo} \\[-1mm]  
                       &     & \quad\SEEA f_2(1270) + \pi^0  \NN \\[-1mm]
                       &     & \quad\quad\quad\SEEA 2\pi^0  \NN
\\
   \ppb(J^{PC}=2^{++}) & \to & f_2(1270) + 2\pi^0  \LL{f2popo} \\[-1mm]  
                       &     & \quad\SEEA 2\pi^0  \NN
\\
   \ppb(J^{PC}=2^{++}) & \to & a_2(1660) + \pi^0  \LL{a2popo} \\[-1mm]  
                       &     & \quad\SEEA f_2(1270) + \pi^0  \NN \\[-1mm]
                       &     & \quad\quad\quad\SEEA 2\pi^0  \NN
\\
   \ppb(J^{PC}=2^{++}) & \to & \pi_2(1670) + \pi^0  \LL{pi2po} \\[-1mm]  
                       &     & \quad\SEEA f_2(1270) + \pi^0 \NN \\[-1mm]
                       &     & \quad\quad\quad\SEEA 2\pi^0  \NN
\\
   \ppb(J^{PC}=2^{++}) & \to & \pi(1300) + \pi^0  \LL{pi13po} \\[-1mm]  
                       &     & \quad\SEEA \sigma + \pi^0 \NN \\[-1mm]
                       &     & \quad\quad\quad\SEEA 2\pi^0  \NN
\\
   \ppb(J^{PC}=2^{++}) & \to & \sigma + f_0  \LL{j2sigf0} \\[-1mm]  
                       &     & \quad\quad\quad\to  4\pi^0  \NN
\\
   \ppb(J^{PC}=2^{++}) & \to & \sigma + \sigma  \LL{j2sigsig} \\[-1mm]  
                       &     & \quad\quad\quad\to  4\pi^0  \NN
\\
   \ppb(J^{PC}=0^{++}) & \to & \sigma + \sigma  \LL{sigsig} \\[-1mm]  
                       &     & \quad\quad\quad\to  4\pi^0  \NN
\\
   \ppb(J^{PC}=0^{++}) & \to & \sigma + f_0  \LL{sigf0} \\[-1mm]  
                       &     & \quad\quad\quad\to  4\pi^0  \NN
\EA
The details of the calculations are given in Appendix~\ref{APA}.

\begin{figure}[htbp]
\centering 
\mbox{(a) \hspace*{70mm} (b)}
\\[0\baselineskip]
\mbox{\hspace*{-10mm}
  \mbox{\epsfysize=7cm\epsffile{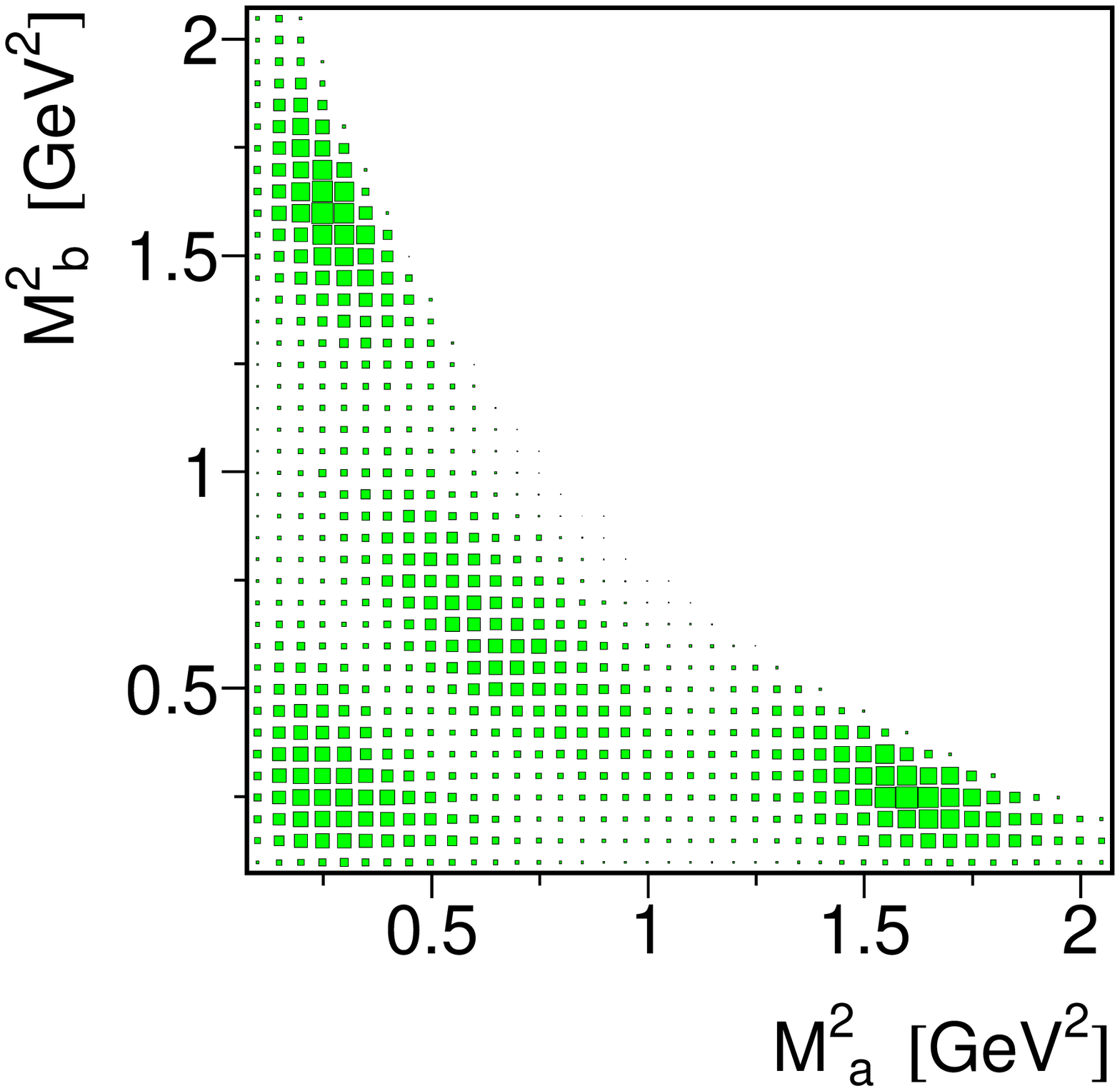}}        
  \mbox{\epsfysize=7cm\epsffile{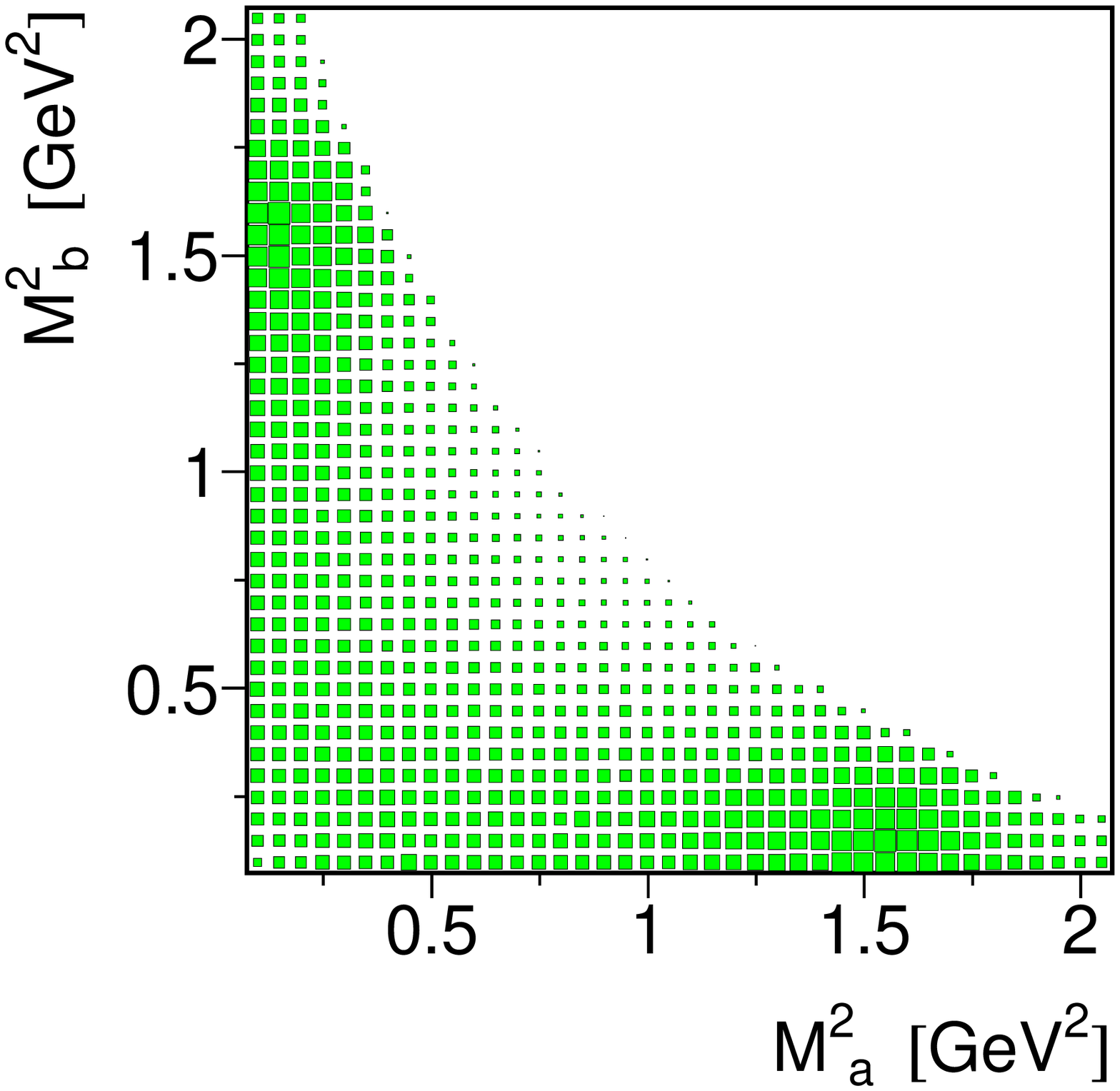}} 
}
\\[0\baselineskip]
\mbox{(c) \hspace*{70mm} (d)}
\\[0\baselineskip]
\mbox{\hspace*{-10mm}
  \mbox{\epsfysize=7cm\epsffile{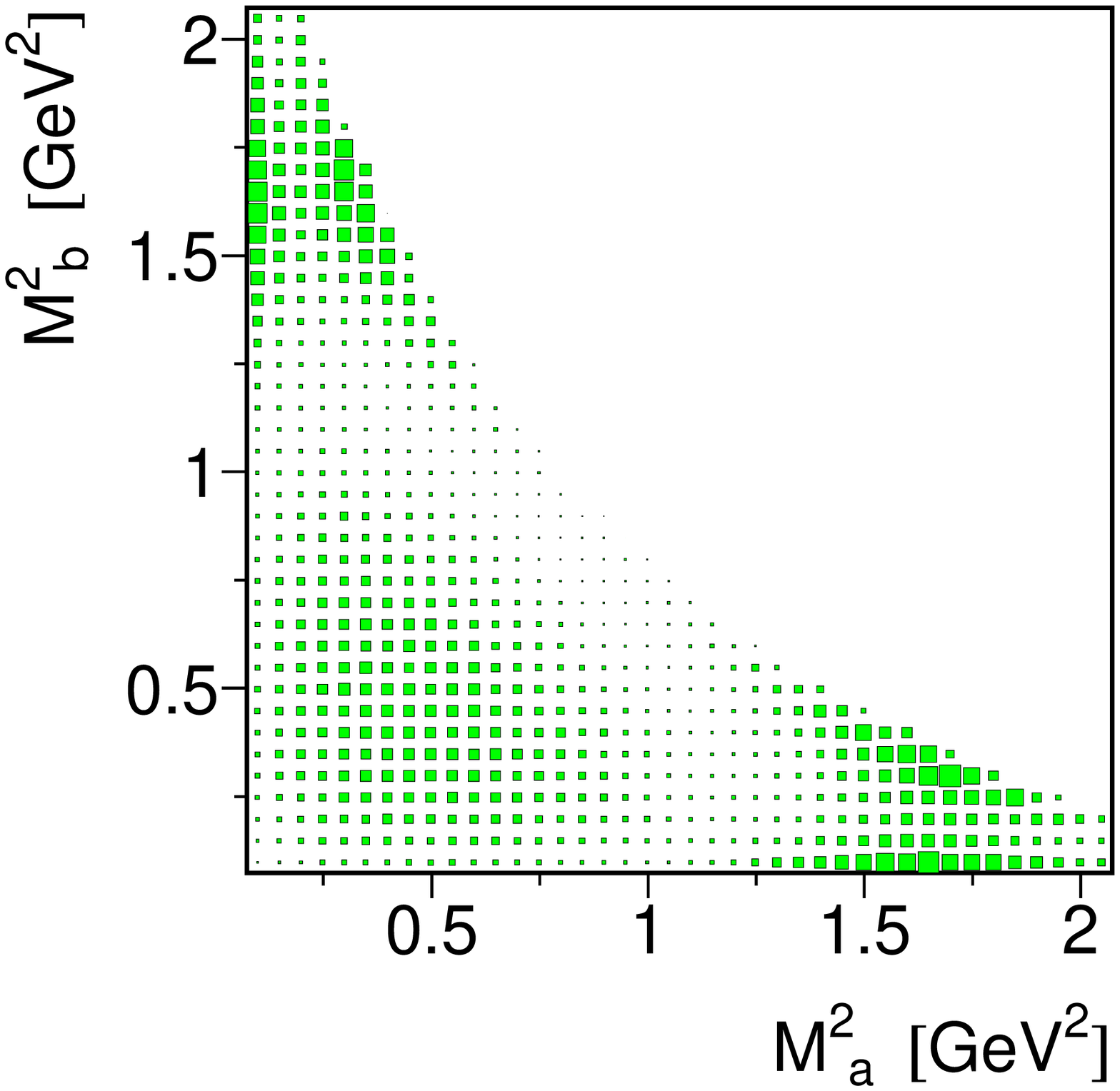}} 
  \mbox{\epsfysize=7cm\epsffile{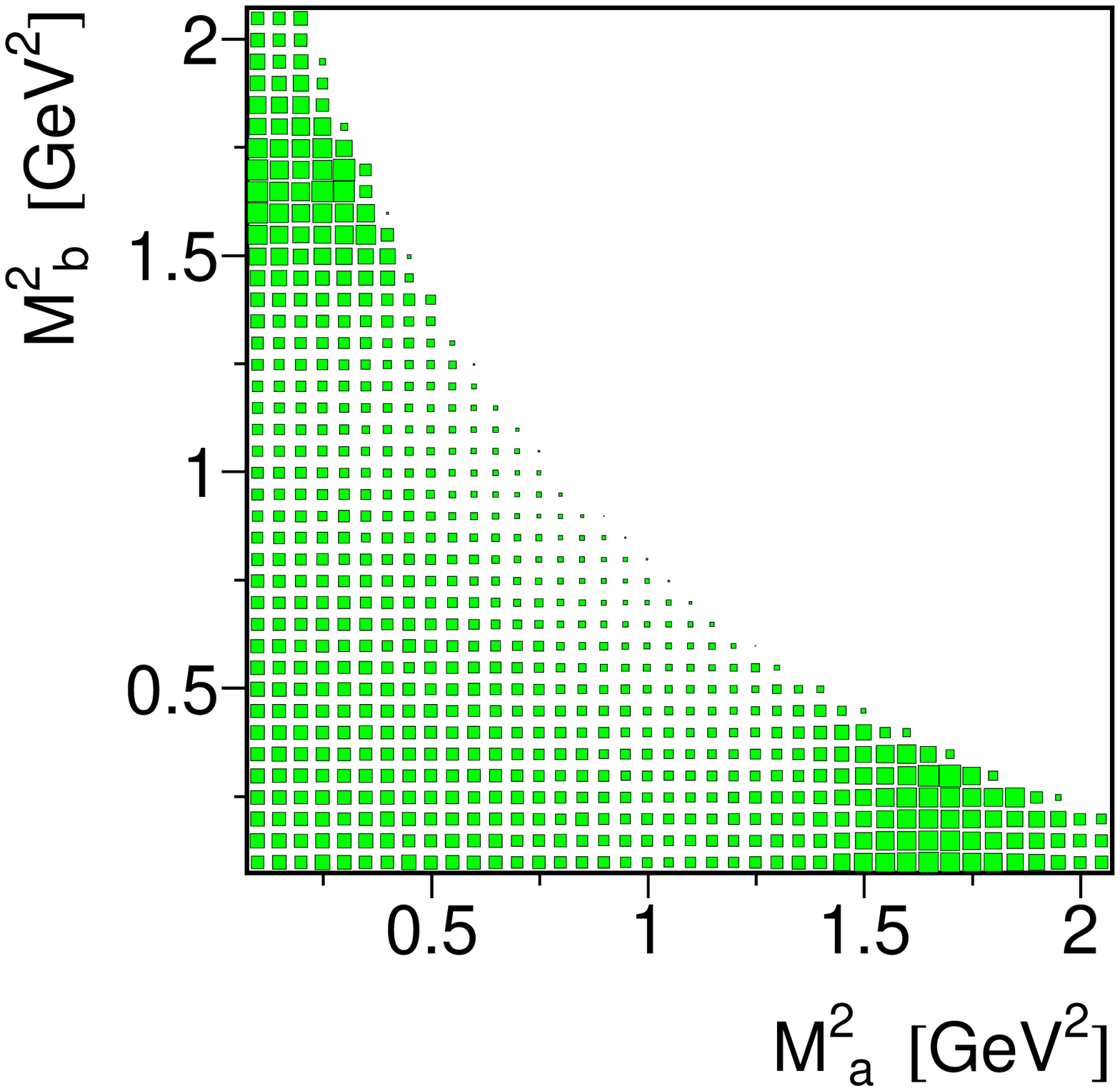}} 
}
\\[0\baselineskip] 
\caption{
 The double-differential density $\RHO(\Mooa,\Moob)$ calculated for 
 different reaction mechanisms:
 (a) $\ppb(J^{PC}=0^{-+}) \to  a_2(1660)_{\to f_2(1270)+\pi^0} + \pi^0$,   
 (b) $\ppb(J^{PC}=2^{++}) \to  f_2(1270) + 2\pi^0$,   
 (c) $\ppb(J^{PC}=2^{++}) \to  a_2(1660)_{\to f_2(1270)+\pi^0} + \pi^0$,   
 (d) $\ppb(J^{PC}=2^{++}) \to  \pi_2(1670)_{\to f_2(1270)+\pi^0} + \pi^0$.    
}
\label{DDXsep}
\end{figure}

\begin{figure}[htbp]
\centering 
\mbox{(a) \hspace*{70mm} (b)}
\\[0\baselineskip]
\mbox{\hspace*{-10mm}
\mbox{\epsfysize=7cm\epsffile{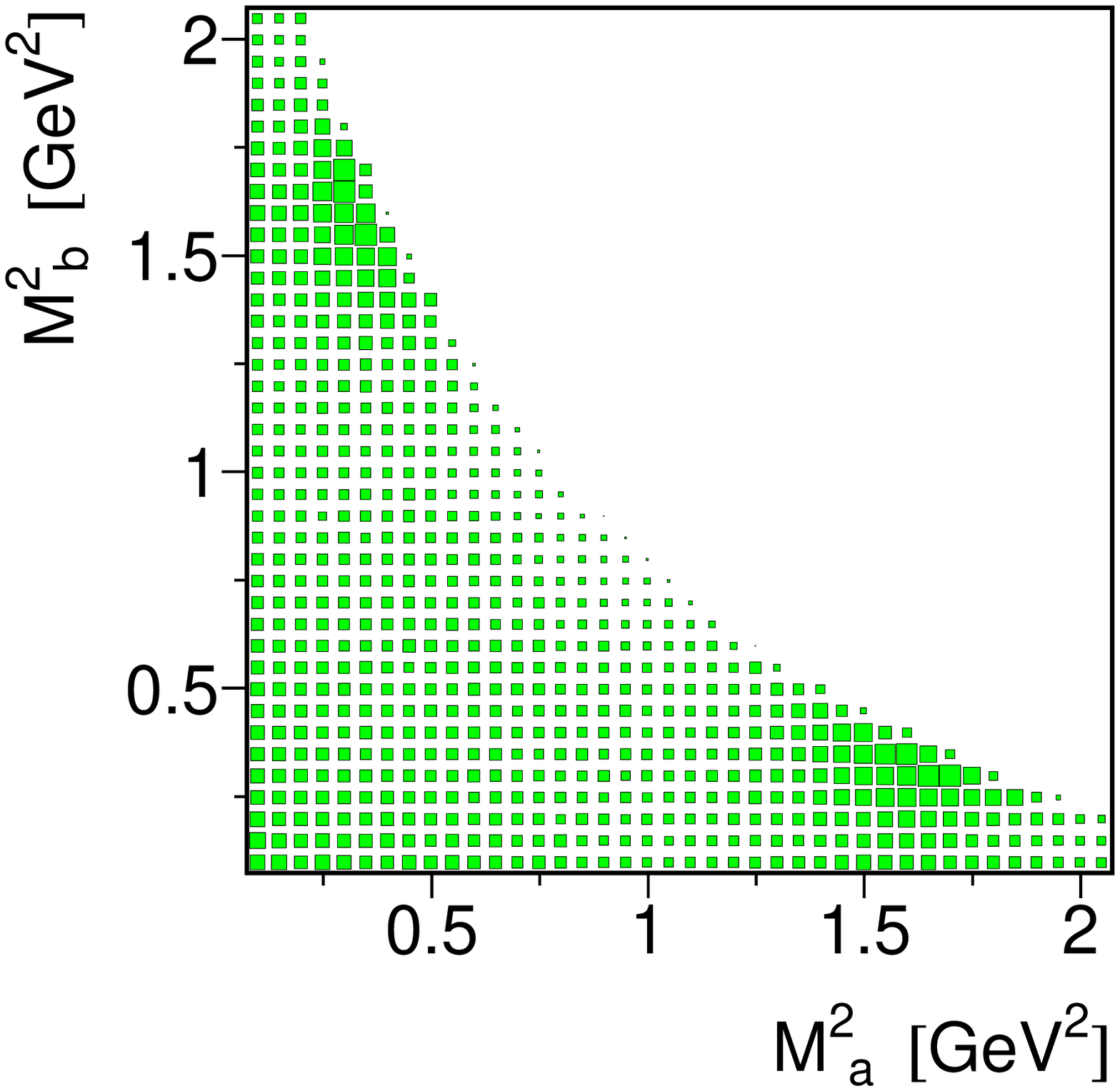}} 
\mbox{\epsfysize=7cm\epsffile{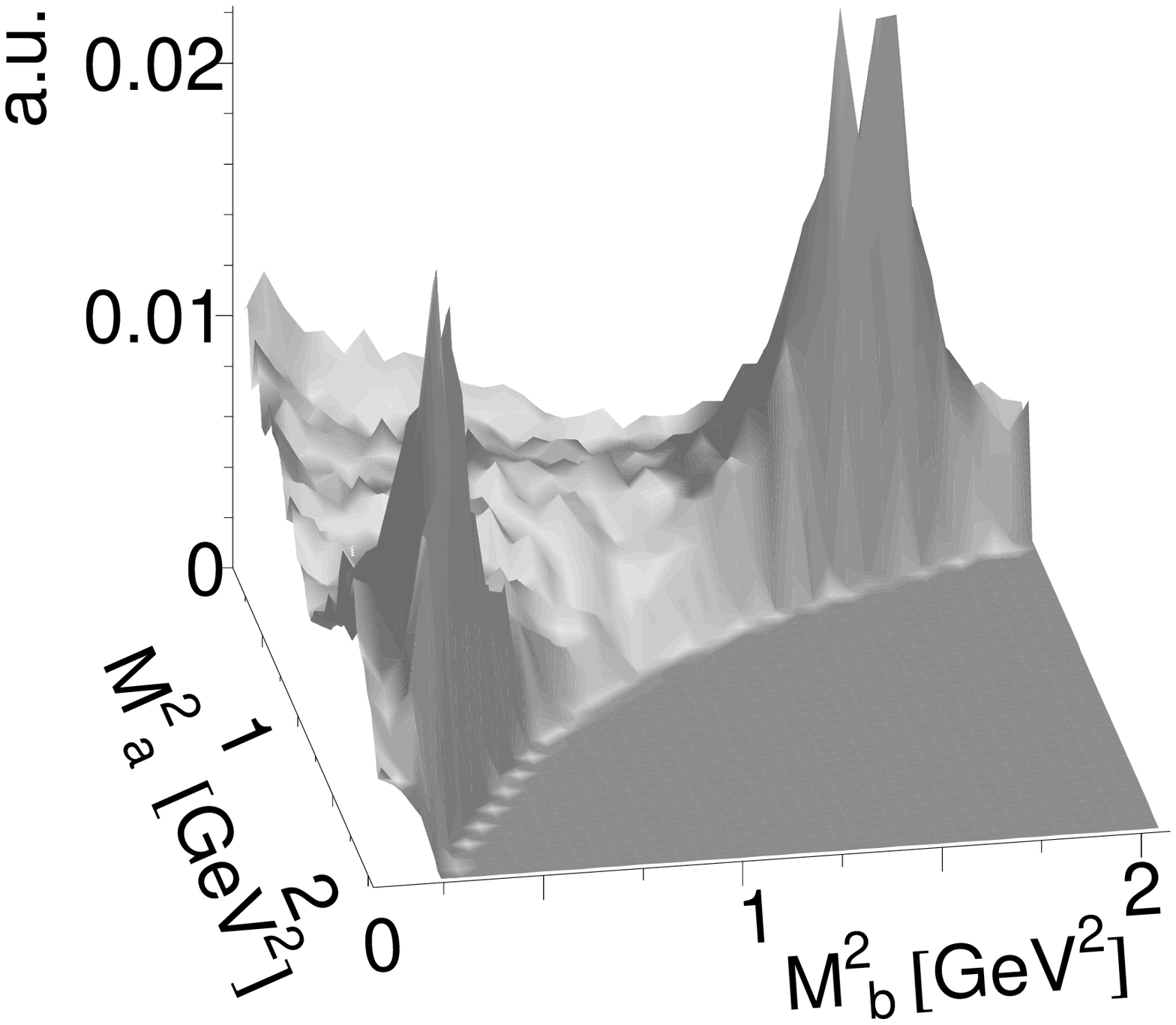}} 
}
\caption{
 The double-differential density $\RHO(\Mooa,\Moob)$ calculated for 
 a sum of the mechanisms (\protect\ref{f2popo}-\protect\ref{pi2po}) 
 as discussed in the text. 
}
\label{DDXmix}
\end{figure}

As a first step, we inspect the double--differential densities corresponding 
to each individual mechanism in order to demonstrate the effects expected
if one of these mechanisms would be dominant.  
In reality, a coherent superposition of several mechanisms is required 
to describe the data \cite{Ko98}, therefore, as a second step, 
we investigate the double-differential distributions corresponding to 
such superpositions.   
Our goal is to find out which mechanisms are required to describe 
the global features of the differential mass distributions. 
Once these features are established we also get a prediction for the 
double differential distributions near the origin from these ordinary 
resonance mechanisms without introducing any HBT-like stochastic correlations.  

  The double--differential densities calculated for the mechanisms 
(\ref{j0ma2popo},\ref{f2popo},\ref{a2popo},\ref{pi2po}) are shown in 
Figs.\ref{DDXsep}(a,b,c,d). 

By varying the relative weights and phases of the individual mechanisms 
we have found that the superpositions of the three mechanisms 
(\ref{f2popo},\ref{a2popo}, and \ref{pi2po}) 
are sufficient to obtain a qualitative agreement with the data:  
\newcommand{\e}[1]{{\mathrm e}^{#1}}
\BA
     T & = &  c_1 T_{f_2(1270)+2\pi^0} 
             +c_2 \e{i\phi_2}\; T_{a_2(1660)+\pi^0} 
             +c_3 \e{i\phi_3}\; T_{\pi_2(1670)+\pi^0} 
\EA  
In fact, the main features of the data can be reproduced 
even with the two mechanisms (\ref{a2popo}) and (\ref{pi2po}) 
(note that they include the contribution from the lighter meson 
$f_2(1270)$, see Figs.\ref{Diags}(b,c)).  
Figure~\ref{DDXmix} shows an example corresponding to a superposition of 
the mechanisms (\ref{a2popo}) and (\ref{pi2po}) with   
$c_1=0$, $c_2=0.45$, $c_3=0.91$, $\phi_2=0$, $\phi_3=3\pi/2$ 
(all individual amplitudes are normalized to the same total yield of $4\pi^0$). 
We are not aiming at an optimal fit to the data in the present context, 
but Fig.~\ref{DDXmix} illustrates that the observed resonance peaks are 
consistent with a moderate enhancement near the origin.  
Compare this figure with the experimental distribution of figure~\ref{rg40}.

\section{Conclusion}
\label{CONCL}

For the annihilation reaction $\ppb \rightarrow 4\pi^0$ at rest, 
the BE correlations between two neutral pions have been studied
for the first time with full event reconstruction.
The inclusive $2\pi^0$ correlation shows an unexpected dip at small
relative momentum, in contrast to the inclusive pair correlation 
in the charged channel $\ppb \rightarrow 2\pi^+2\pi^-$ which shows 
a weak enhancement~\cite{CPLEAR97}.

For the more sensitive and less model-dependent double-differential
distribution, a weak peaking is observed near the origin 
(small invariant masses of the two neutral pion pairs).  This 
is similar to though less pronounced than in the $2\pi^+2\pi^-$ channel.

The data of the single- and double-differential 
$2\pi^0$ correlations were compared to model calculations.  
The role of the $\rho$ meson dominating the $2\pi^+2\pi^-$ dynamics is 
replaced by the $a_2$, $\pi_2$, and $f_2$ resonances in the $4\pi^0$ case.
Resonance production explains the fairly different spectra of the 
two reactions.  It is found that with an adequate
choice of reaction amplitudes the dynamical model describes the main 
features of the single- and double-differential distributions, 
including a slight enhancement near the origin.

From this result, together with the absence of a correlation signal 
in the inclusive $2\pi^0$ correlation, we conclude that our analysis 
does not favour the interpretation of pion correlation signals in 
terms of an HBT-type model with stochastic pion emission phases.

\section{Aknowledgements}

The authors are very grateful to the Crystal Barrel Collaboration for
making available the data.  In this connection we would like to thank in
particular Ulrich Wiedner and \v{C}rtomir Zupan\v{c}i\v{c}. 
We had stimulating discussions on aspects of pion correlations with 
Eberhard Klempt and Ulrike Thoma.  
One of us (MPL) would like to thank the TRIUMF theory
group for hospitality during completion of this paper.

\appendix
\section{Appendix} \LL{APA}

\begin{figure}[htbp]
\centering 
\mbox{(a) \hspace*{50mm} (b)  \hspace*{50mm} (c) }
\\[0\baselineskip]
\mbox{\hspace*{-10mm}
\mbox{\epsfysize=50mm\epsffile{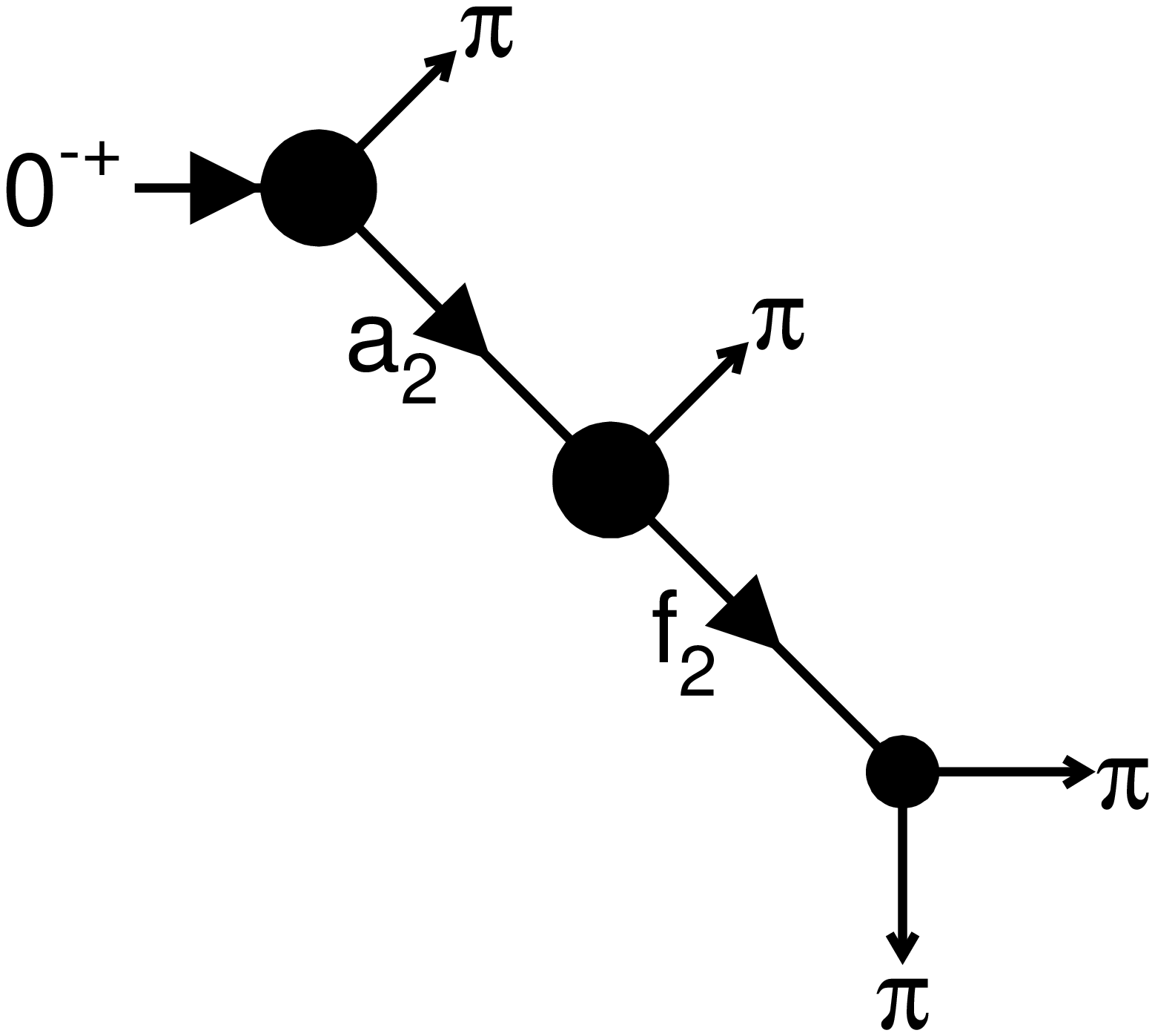}} 
\mbox{\epsfysize=50mm\epsffile{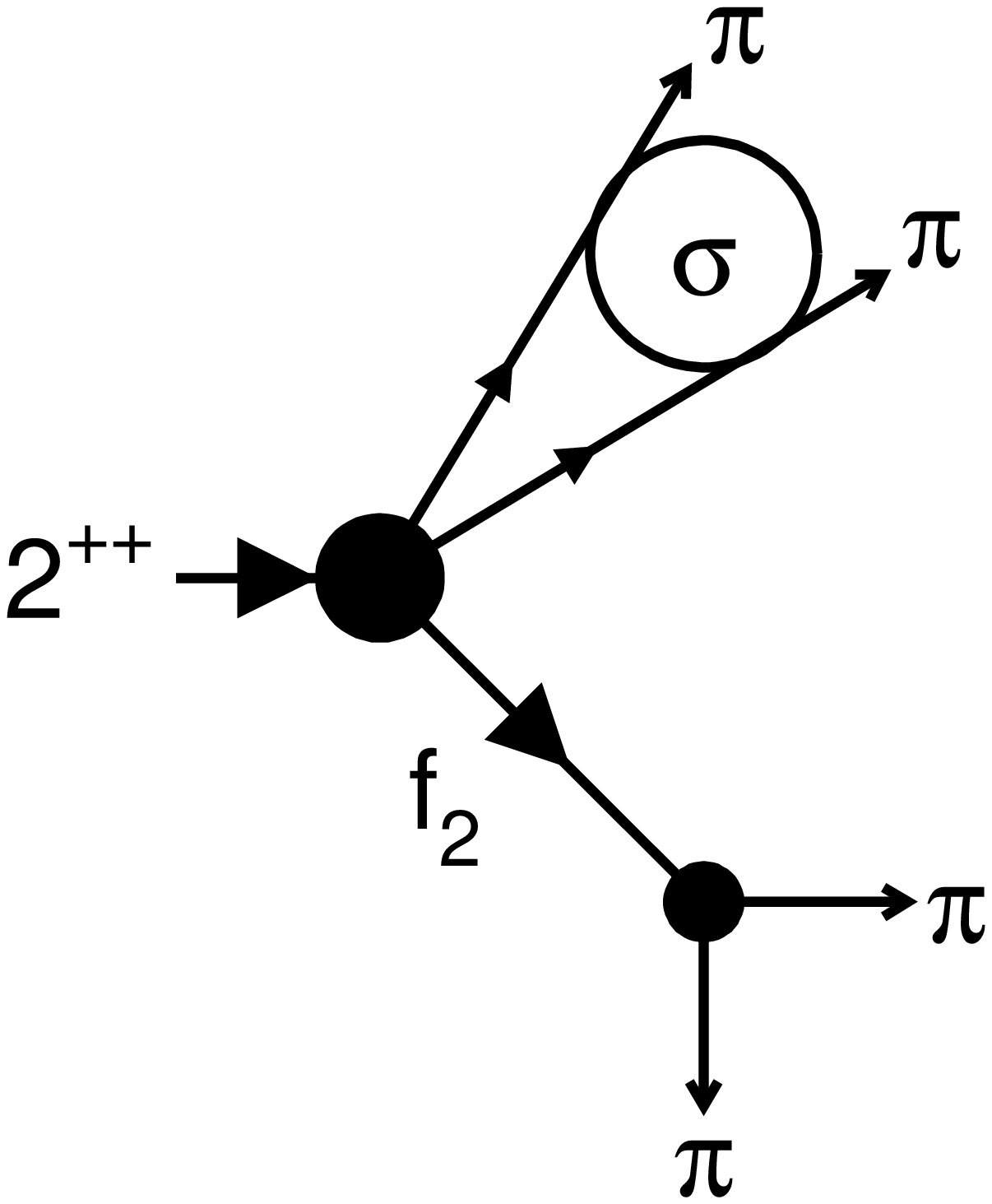}} 
\mbox{\epsfysize=50mm\epsffile{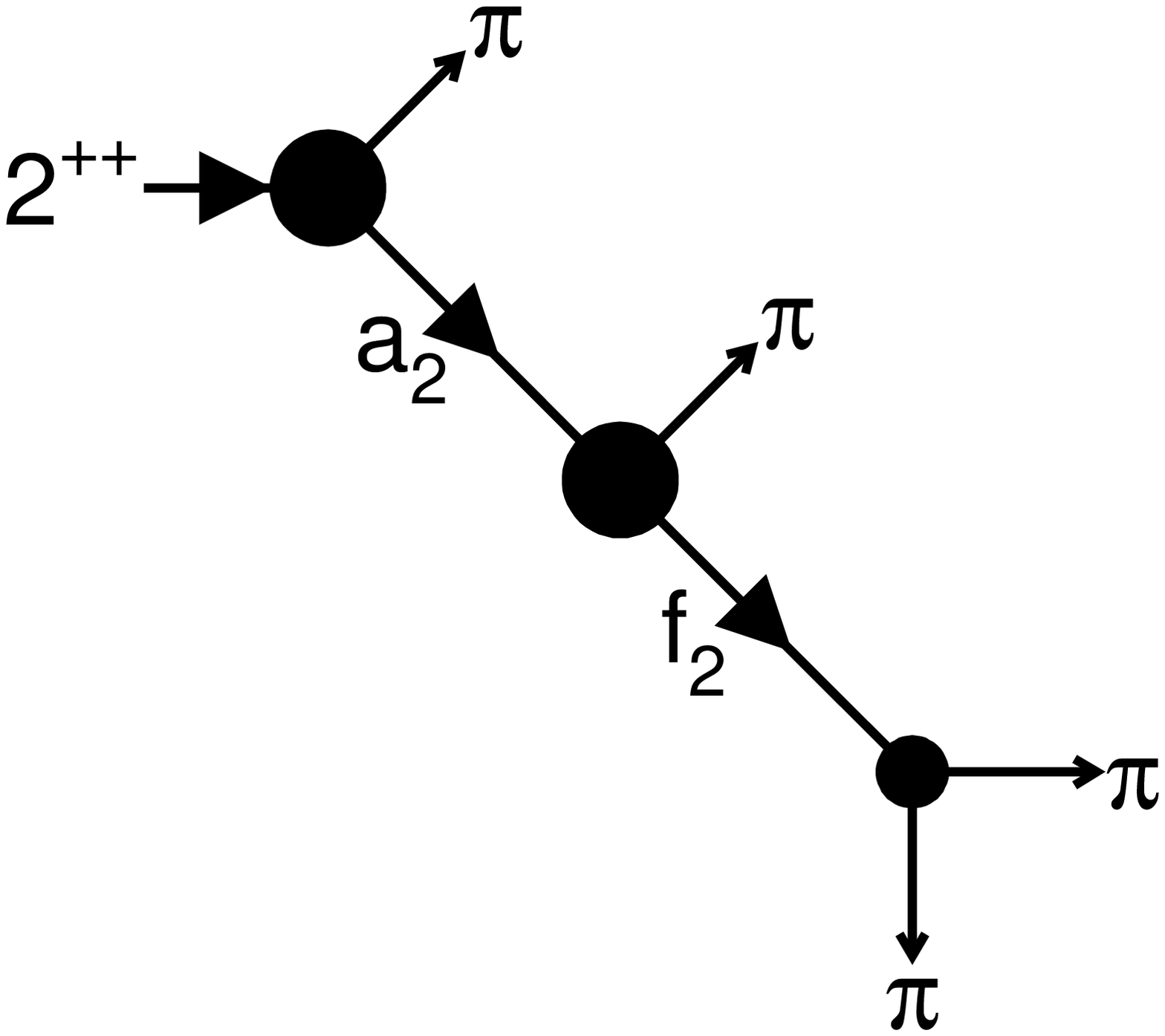}} 
}
\\[0\baselineskip]
\mbox{(d) \hspace*{50mm} (e)  \hspace*{50mm} (f) }
\\[0\baselineskip]
\mbox{\hspace*{-10mm}
\mbox{\epsfysize=50mm\epsffile{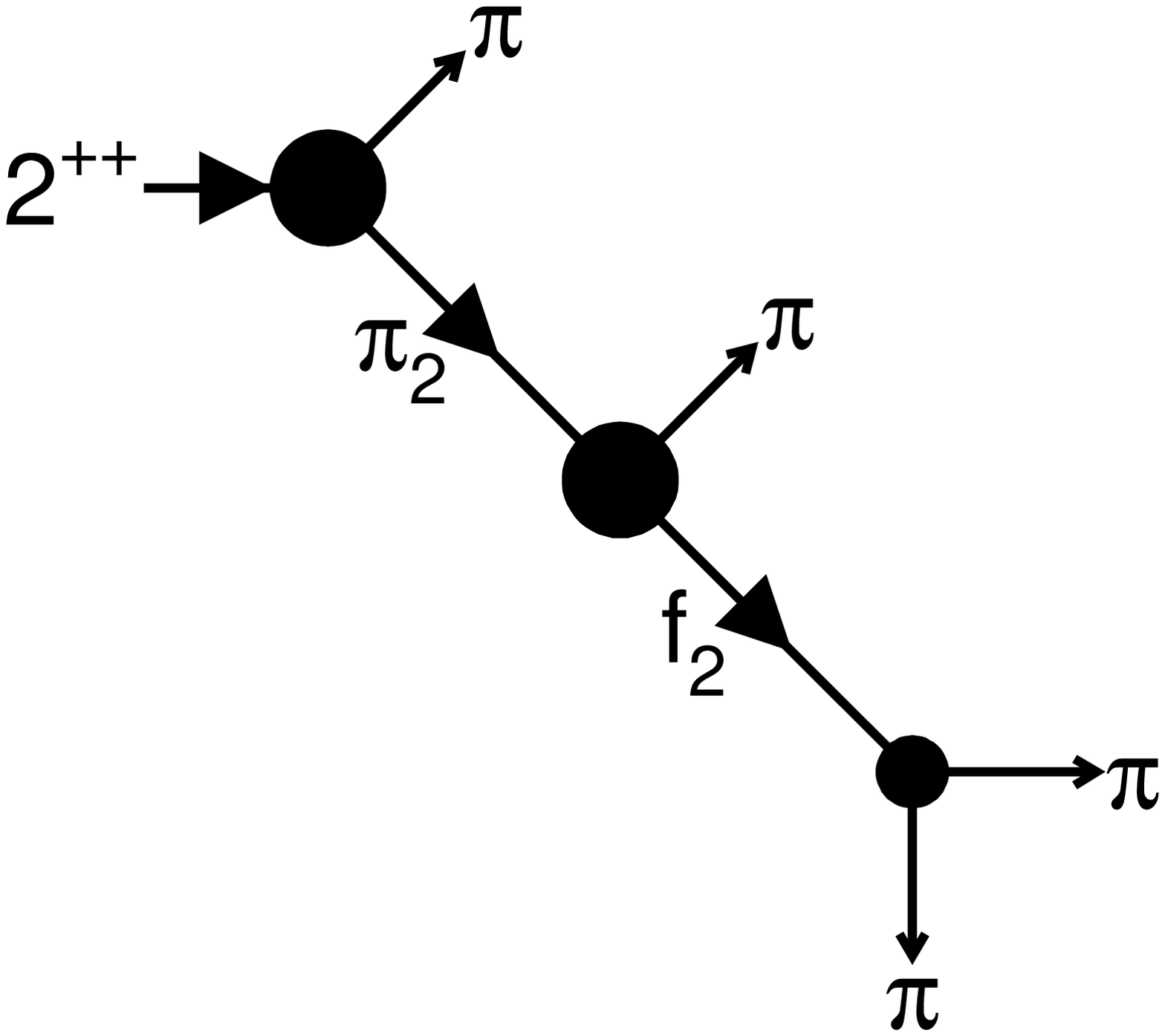}} 
\mbox{\epsfysize=50mm\epsffile{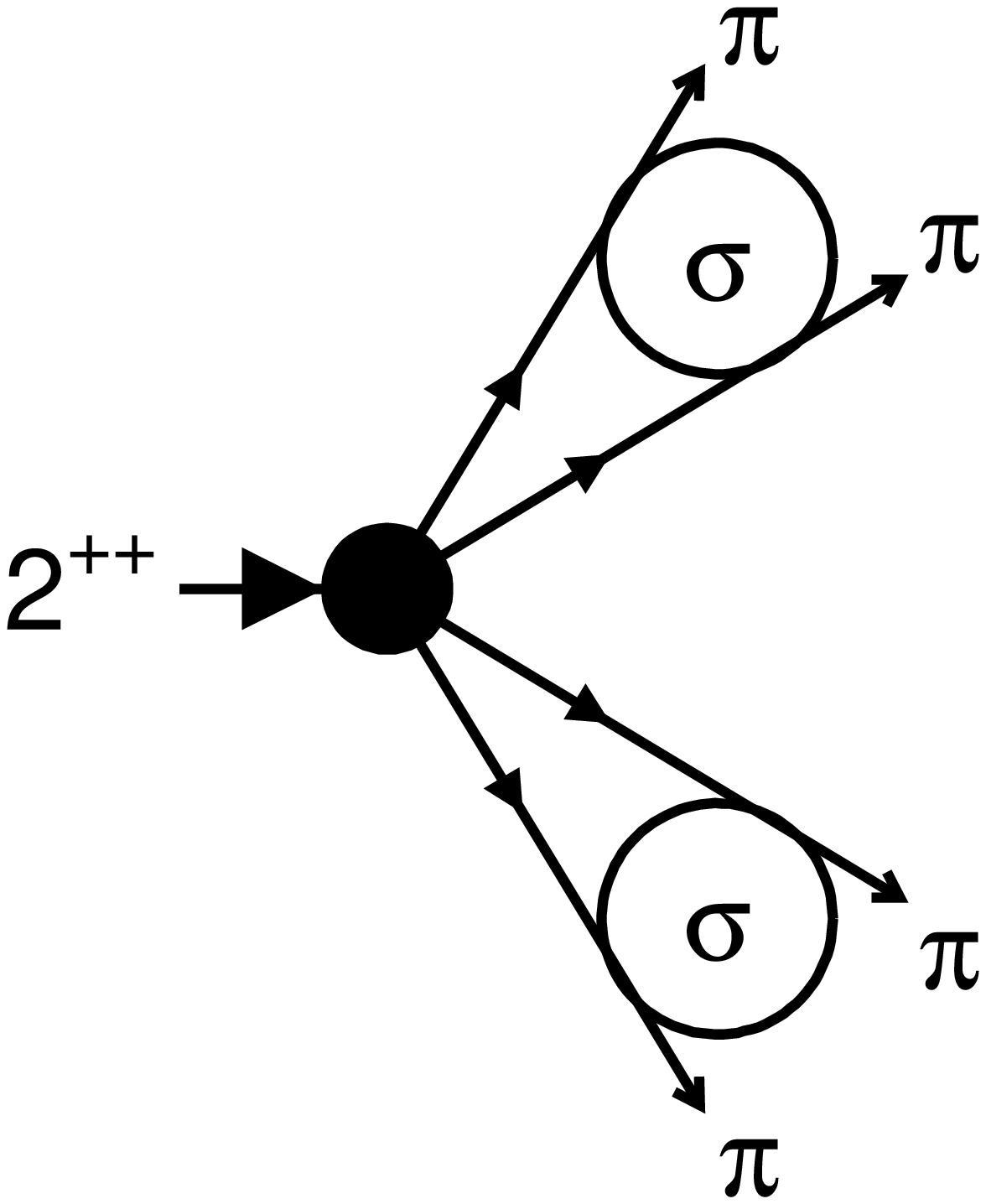}} 
\mbox{\epsfysize=50mm\epsffile{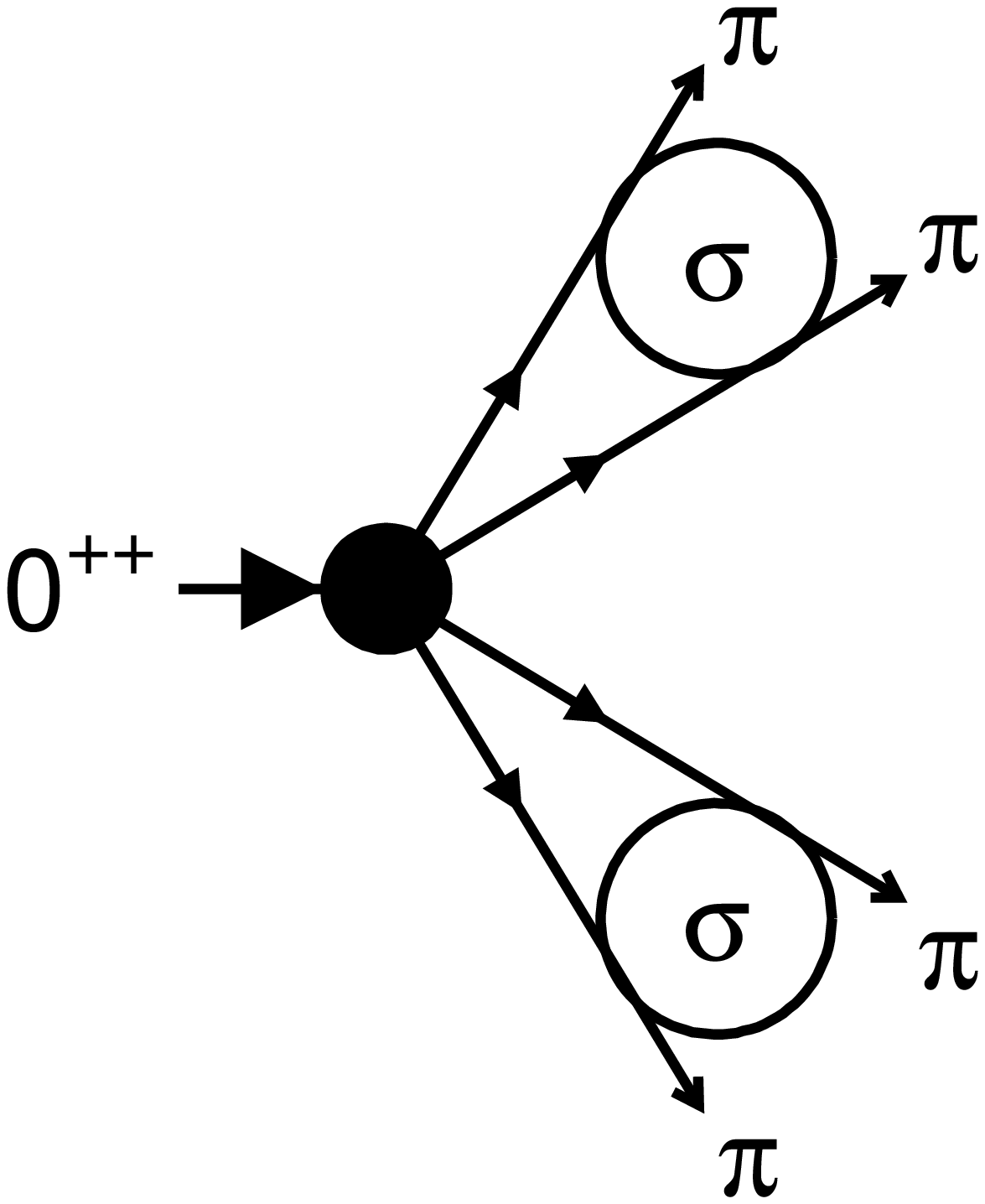}} 
}
\caption{
   The resonance mechanisms. 
}
\label{Diags}
\end{figure}

The amplitudes corresponding to the resonance mechanisms 
(\ref{f2popo},\ref{a2popo},\ref{pi2po},\ref{j2sigsig},\ref{j0ma2popo},\ref{sigsig}) 
are given by the diagrams in figures~\ref{Diags}.  
The calculations are straightforward, so only some major elements are 
outlined here. 
The $\ppb$ vertices are taken with a minimum number of derivatives. 
The vertex $\ppb(J^{PC}=2^{++})\to a_2\pi$ has the form
\BA
  V_{\ppb(J^{PC}=2^{++})\to a_2\pi} & = &
    B_{\mu\nu} g^{\nu\gamma} T_{\gamma\delta} 
    \epsilon^{\alpha\beta\mu\delta} (p_{\pi})_{\alpha} (p_{a_2})_{\beta} 
\EA
where $B_{\mu\nu}$ is the polarization tensor of the $\ppb(J^{PC}=2^{++})$ 
state and $T_{\gamma\delta}$ is the polarization tensor of the $a_2$, 
and $p_{\pi}$ and $p_{a_2}$ are the momenta of the corresponding particles.    
The $a_2\to f_2 \pi$ vertex has a similar structure, and 
the $\ppb(J^{PC}=2^{++})\to \pi_2\pi$ is a trivial $S$-wave vertex.   
The $f_2\pi\pi$ vertex has the form
\BA
  V_{f_2\pi\pi} & = &
  g_{f_2\pi\pi} T_{\mu\nu} q^{\mu} q^{\nu}  F_{f_2\pi\pi}(m_{\pi\pi}^2) 
\EA
where $g_{f_2\pi\pi}$ is the coupling constant, 
$T_{\mu\nu}$ is the polarization tensor of the $f_2$,   
$q = p_1 - p_2$, 
$p_n$ are the pion momenta, and
$F_{f_2\pi\pi}(m_{\pi\pi}^2)$ is the form factor depending on the invariant
mass of the $\pi\pi$ system.  

The propagators corresponding to the tensor particles are given by
the formula
\BA
    G(p) & = &
    \frac{ \frac{1}{2}(\Pi^{\mu\mu'}\Pi^{\nu\nu'}+\Pi^{\nu\mu'}\Pi^{\mu\nu'}) -
           \frac{1}{3}(\Pi^{\mu\nu}\Pi^{\mu'\nu'}) }
         {p^2-m_0^2-{\cal M}(p^2)}
\\
    \Pi^{\mu\nu} & = & g^{\mu\nu}-p^{\mu}p^{\nu}/p^2 
\EA
where $m_0$ is the bare mass and ${\cal M}(p^2)$ is the mass operator.  
The mass operator for the $f_2$ meson corresponding to the $\pi\pi$ loop 
is defined by the dispersion integral:
\BA
   {\cal M}(s) & = & \frac{1}{2\pi}
   \int_{4m_{\pi}^2}^{\infty} \frac{\Gamma(s')}{s-s'} ds' 
\\
   \Gamma(s) & = & \frac{g_1^2 |F_{f_2\pi\pi}(s)|^2 k^5}{60\pi s} 
\EA
where $k=k(s)=\sqrt{s/4 - m_{\pi}^2}$ is the relative momentum in the $\pi\pi$ system 
and the dipole form factor   
$F_{f_2\pi\pi}(s)=(1+k(s)^2/\nu^2)^{-2}$ is used with $\nu=1\;$GeV.  
The parameters $m_0$ and $g_{f_2\pi\pi}$ are defined by the mass and the 
width of the $f_2$ meson, with the other decay channels being approximated by 
a constant width.     
The propagators of $a_2$ and $\pi_2$ are constructed in a similar way. 
The $\sigma$ block in Fig.\ref{Diags}a denotes the full Green function
of the $\pi\pi$ system in the scalar-isoscalar channel; it was taken from 
the coupled channel model \cite{LMZ}. 
The components above are combined to construct the amplitudes for 
the resonance channels concerned; since the procedure is straightforward, we 
skip the resulting formulas involving lengthy tensor structures.


\end{document}